\documentclass[11pt]{article}
\usepackage{aaspp4}
\slugcomment{Accepted for publication in \textit{The Astrophysical Journal}}

\lefthead{Clegg, Fey, \& Lazio}
\righthead{Gaussian Plasma Lenses}

\begin{document}

\title{The Gaussian Plasma Lens in Astrophysics.  Refraction}

\author{Andrew~W.~Clegg\altaffilmark{1}}
\affil{Naval Research Laboratory, Code 7213, Washington, DC  20375-5351\\
	clegg@intrepid.nrl.navy.mil}
\altaffiltext{1}{Current address: Millimeter Wave Report, P.O.~Box~4749,
	Arlington, VA  22204-0749; mmwr@msn.com}

\author{Alan~L.~Fey}
\affil{United States Naval Observatory, Code EO\\ 
	3450 Massachusetts Ave.~NW, Washington, DC  20392-5420\\
	afey@alf.usno.navy.mil}

\and

\author{T.~Joseph~W.~Lazio\altaffilmark{2}}
\affil{Naval Research Laboratory, Code 7210, Washington, DC  20375-5351\\
	lazio@rsd.nrl.navy.mil}
\altaffiltext{2}{NRC-NRL Research Associate}

\authoraddr{Address for correspondence:
	T. Joseph W. Lazio
	NRL, Code 7210
	Washington, DC  20375-5351}	

\begin{abstract}
We present the geometrical optics for refraction of a distant
background radio source by an interstellar plasma lens, with specific
application to a lens with a Gaussian profile of free electron column
density.  The refractive properties of the lens are specified
completely by a dimensionless parameter, $\alpha$, which is a function
of the wavelength of observation, the free electron column density
through the lens, the lens-observer distance, and the diameter of the
lens transverse to the line of sight.  A lens passing between the
observer and a background source, due to the relative motions of the
observer, lens, and source, produces modulations in the light curve of
the background source.  Because plasma lenses are diverging, the light
curve displays a minimum in the background source's flux density,
formed when the lens is on-axis, surrounded by enhancements above the
nominal (unlensed) flux density.  The exact form of the light curve
depends only upon the parameter $\alpha$ and the relative angular
sizes of the source and lens as seen by the observer.  Other effects
due to lensing include the formation of caustic surfaces, upon which
the apparent brightness of the background source becomes very large;
the possible creation of multiple images of the background source; and
angular position wander of the background source.  If caustics are
formed, the separation of the outer caustics can be used to constrain
$\alpha$, while the separation of the inner caustics can constrain the
size of the lens.  We apply our analysis to two sources which have
undergone extreme scattering events: 0954$+$654, a source for which we
can identify multiple caustics in its light curve, and 1741$-$038, for
which polarization observations were obtained during and after the
scattering event.  We find general agreement between modelled and
observed light curves at 2.25~GHz, but poor agreement at 8.1~GHz.  The
discrepancies between the modelled and observed light curves may
result from some combination of substructure within the lens, an
anisotropic lens shape, a lens which only grazes the source rather than
passing completely over it, or unresolved substructure within the
extragalactic sources.  Our analysis also allows us to place
constraints on the physical characteristics of the lens.  The inferred
properties of the lens responsible for the scattering event toward
0954$+$654 (1741$-$038) are that it was 0.38~AU (0.065~AU) in
diameter, with a peak column density of 0.24~pc~cm${}^{-3}$
($10^{-4}$~pc~cm${}^{-3}$), an electron density within the lens of
$10^5$~cm${}^{-3}$ (300~cm${}^{-3}$), and a mass of $6.5 \times
10^{-14}$~M${}_{\sun}$ ($10^{-18}$~M${}_{\sun}$).  The angular
position wander caused by the lens was 250~mas (0.4~mas) at 2.25~GHz.
In the case of 1741$-$038, we can place an upper limit of only 100~mG
on the magnetic field within the lens.
\end{abstract}


\section{Introduction}\label{intro}

Strong refraction by interstellar electron density irregularities has
been invoked to explain several radio astronomical phenomena.  Fiedler
et al.~(1987, 1994a) and Clegg, Fey, \& Fiedler~(1996) reported a
number of extreme scattering events (ESEs)---dramatic changes in the
2.25~GHz flux density---in the light curves of several extragalactic
radio sources observed during the course of a 2.25 and~8.1~GHz
monitoring program.  The changes typically lasted several weeks or
months and were occasionally accompanied by strong variations at
8.1~GHz.  Because of the simultaneity of the events at different
wavelengths, the time scales of the events, and speed-of-light travel
time arguments, Fiedler et al.~(1987) and Romani, Blandford, \&
Cordes~(1987) concluded that the flux density variations were likely
due to strong scattering by the Galactic interstellar medium (ISM).  A
similar dramatic change in the flux density, and times of arrival, has
since been observed during a timing program of the pulsar
PSR~B1937$+$21 (\cite{cblbadd93}).

Interstellar refraction may also be responsible for episodes of
multiple imaging and fringing observed in various pulsars' dynamic
spectra (Hewish, Wolszczan, \& Graham~1985; \cite{wc87}; Gupta, Rickett,
\& Lyne~1988; Clegg, Fiedler, \& Cordes~1993).  Dynamic spectra record
the flux density of the pulsar as a function of time and frequency.
Because a pulsar is an angularly compact coherent emitter, strong
interstellar refraction may produce multiple images of the pulsar.
The multiple pulsar images can then interfere with one another,
producing fringes in the time and frequency domains which are visible
in the dynamic spectra as alternating bright and dark bands modulating
the diffractive scintillation pattern of the pulsar.

The characteristics of the medium responsible for strong interstellar
refraction are not clear.  For example, it is unknown if the
refraction responsible for such phenomena as ESEs and fringing in
pulsar dynamic spectra is the result of a localized large-amplitude
electron density enhancement (a ``plasma lens'') located somewhere
between the observer and source, or whether the refraction is the
result of an extended turbulent medium with a large number of electron
density fluctuations distributed along the line of sight.  There are
arguments that favor both models: Clegg, Chernoff, \& Cordes~(1988)
showed that expanding shock fronts can naturally create discrete
plasma lenses, while the general ISM is widely believed to have an
extended, turbulent distribution of electron density fluctuations
(\cite{r90}).

To date, refraction phenomena which have been interpreted in terms of
the discrete lens model have been analyzed using back-of-the-envelope
calculations for the effects of the lens (e.g., \cite{cfc93}).  These
calculations have been useful for deriving approximate characteristics
of the background radio source and the lens itself.  The purpose of
this paper is to venture beyond the back-of-the-envelope calculations
and to present a detailed, quantitative description of the optics
properties of an interstellar plasma lens.  The motivation is to
present the optics of interstellar plasma lenses in a convenient
format that allows the application to a wide variety of problems.

In this paper, the analysis is limited to the refractive properties
of a one-dimensional plasma lens.  Results are applied to the specific
case of a lens with a Gaussian distribution of free electron density
across the lens plane.  The Gaussian lens is
a convenient form for describing a plasma irregularity confined to
a characteristic size, and whose influence on the background wave front
is negligible outside of this characteristic size.  

The plan of this paper is the following. In \S\ref{sec:geo_optics} we
present a general description of the geometrical optics of a plasma
lens, in \S\ref{sec:gausslens} we specialize the application to the
case of a Gaussian lens form, in \S\ref{sec:lt_curve} we discuss the
light curves expected from a Gaussian plasma lens passing in front of
both point and extended sources, in \S\ref{sec:caustics} we discuss
the formation of caustics and their utility for extracting information
about the lens from the light curve, in \S\ref{sec:ang_dis_sec} we
discuss the expected angular position wander caused by a Gaussian
lens, and in \S\ref{sec:ese} we apply our model to the extreme
scattering events seen toward the sources 0954$+$658 and 1741$-$038.
We summarize our conclusions in the last section.

\section{Geometrical Optics of a Plasma Lens}\label{sec:geo_optics}

The radio frequency index of refraction $n_{\mathrm{r}}$ in an
unmagnetized plasma is
\begin{eqnarray}
n_{\mathrm{r}}^2(\omega) & = & 1 - \omega_{\mathrm{e}}^2/\omega^2 \nonumber \\
 & = & 1 - \lambda^2 r_{\mathrm{e}} n_{\mathrm{e}} / \pi,
\label{eqn:refindex}
\end{eqnarray}
where $\omega_{\mathrm{e}}^2 = 4\pi n_{\mathrm{e}} e^2 /
m_{\mathrm{e}}$ is the square of the electron plasma frequency; $e$,
$m_{\mathrm{e}}$, $r_{\mathrm{e}} = e^2/(m_{\mathrm{e}} c^2)$, and
$n_{\mathrm{e}}$ are respectively the electron charge, mass, classical
radius, and local number density, ($\omega_{\mathrm{e}} =
56\,\mathrm{s}^{-1}\sqrt{n_{\mathrm{e}}}$ for $n_{\mathrm{e}}$ in
units of cm${}^{-3}$); $\omega = 2\pi c/\lambda$ is the angular
frequency of the wave; $\lambda$ is its free-space wavelength; and $c$
is the speed of light.

The phase velocity through the lens is $v_\phi = c/n_{\mathrm{r}}$,
which is always greater than $c$.  Compared to the time taken by a
wave in vacuum, the phase will traverse a lens of size $l$ in a time
$\tau = l(1/c - 1/v_\phi)$ less.  In the limit $\omega_{\mathrm{e}}
\ll \omega$, which is always true in the case of radioastronomical
observations, the emergent phase from the lens will be advanced by an
amount
\begin{equation}
\phi = \omega \tau = \lambda r_{\mathrm{e}} N_{\mathrm{e}}
\label{eqn:advance}
\end{equation}
compared to a wave in vacuum, where $N_{\mathrm{e}} = n_{\mathrm{e}}
l$ is the column density through the lens.  Because the phase is
advanced, plasma lenses are \emph{diverging}.

Allow a one-dimensional plasma lens to have a profile of column
density $N_{\mathrm{e}}(x)$; implicit in this discussion of a plasma
``lens'' is the notion that $N_e(x)$ varies across the lens in a
regular fashion.  A plane wave from an infinitely distant point source
is incident on the lens with phase $\phi_0$.  The phase as a function
of $x$ on the emergent side of the screen will be
\begin{equation}
\phi(x) = \phi_0 + \lambda r_{\mathrm{e}} N_{\mathrm{e}}(x).
\label{eqn:phprofile}
\end{equation}
In the geometrical optics limit, rays of energy flux propagate normal
to the surfaces of constant phase.  As $N_{\mathrm{e}}$ varies across
the lens, there will be a slope to the emerging constant phase
surfaces.  The rays are therefore refracted by the lens through an
angle $\theta_{\mathrm{r}}$ (\cite[Chapter~III]{bw80}):
\begin{eqnarray}
\theta_{\mathrm{r}}(x) & = & \frac{\lambda}{2\pi}\frac{d}{dx} \phi(x), \nonumber \\
 & = & \frac{\lambda^2 r_{\mathrm{e}}}{2\pi}\frac{d}{dx} N_{\mathrm{e}}(x).
\label{eqn:ref_ang}
\end{eqnarray}

The observer is a distance $D$ from the lens,
Fig.~\ref{fig:lens_geom}\textit{a}.  The coordinate in the observer's
plane is given by $x^\prime$.  A ray emerging from the plasma lens at
coordinate $x$ is refracted by an angle $\theta_{\mathrm{r}}(x)$ and
strikes the observer's axis at a position
\begin{equation}
x^\prime = x - \theta_{\mathrm{r}}(x) D,
\label{eqn:ray_geom}
\end{equation}
for $\theta_{\mathrm{r}}(x) \ll 1$, which will be the case throughout
this paper.  Multiple solutions $x = x_k$ (with $k = 1$, 2, \dots, $n$)
to equation~(\ref{eqn:ray_geom}) can exist for a given $x^\prime$.  In
such cases, an observer at $x^\prime$ sees $n$ images of the
background source appearing to come from the directions towards $x_1$,
$x_2$, \dots, $x_n$.  The brightness of each image will be enhanced
(or reduced) by some factor $G_k$ due to refractive focusing (or
defocusing).  The factor $G_k$ is determined by the spreading or
focusing of rays due to the gradient in refraction angle across the
lens:
\begin{eqnarray}
G_k & = & \frac{dx}{dx^\prime}\Big\vert_{x=x_k} \nonumber \\
 & = & \left[1-D\frac{d\theta_{\mathrm{r}}}{dx}\right]^{-1}_{x=x_k}.
\label{eqn:rgain}
\end{eqnarray}
where $x_k$, as a solution to equation~(\ref{eqn:ray_geom}), is
ultimately dependent on $x^\prime$, $D$, $\lambda$, and the function
$N_{\mathrm{e}}(x)$, so that $G_k =
G_k[x^\prime,D,\lambda,N_{\mathrm{e}}(x)]$.

The same analysis can be modified slightly to account for an angularly
extended background source.  In this case, a source of angular extent
$\theta_{\mathrm{s}}$ is considered to be the superposition of point
sources distributed over a range of angles $-\theta_{\mathrm{s}}/2 \le
\theta \le +\theta_{\mathrm{s}}/2$.  For an infinitely distant source
the distribution of rays incident upon the lens screen has a range of
incidence angles $-\theta_{\mathrm{s}}/2 \le \theta_{\mathrm{i}} \le
+\theta_{\mathrm{s}}/2$.  The intensity of the rays coming from each
incidence angle $\theta_{\mathrm{i}}$ is weighted by the function
$B(\theta_{\mathrm{i}})$ to account for the brightness distribution
across the source.

For rays incident upon the lens screen with incidence angle
$\theta_{\mathrm{i}}$, we obtain
\begin{equation}
x^\prime = x - [\theta_{\mathrm{r}}(x) + \theta_{\mathrm{i}}]D.
\label{eqn:ray_geom_skew}
\end{equation}
The term $\theta_{\mathrm{i}} D$ in equation~(\ref{eqn:ray_geom_skew})
accounts for the skewness of the incident ray with respect to the lens
axis, Fig.~\ref{fig:lens_geom}\textit{b}.  The gain factor $G_k$ is
now also dependent on $\theta_{\mathrm{i}}$: $G_k =
G_k[x^\prime,D,\theta_{\mathrm{i}},\lambda, N_{\mathrm{e}}(x)]$.

The total observed intensity of the background source is obtained by
integrating the product of the gain factor $G_k$ and the source
brightness distribution $B(\theta)$ over all $\theta$, and summing the
result of this integration over the $n$ images of the background
source:
\begin{equation}
I[x^\prime,D,\lambda,N_{\mathrm{e}}(x)]
 = \sum_{k=1}^n \int_{-\infty}^{+\infty} B(\theta) G_k[x^\prime,D,\lambda,\theta,N_{\mathrm{e}}(x)]\,d\theta.
\label{eqn:total_I}
\end{equation}

\section{The Gaussian Plasma Lens}\label{sec:gausslens}

The basis of our analysis of lensing is studying the behavior of
$I[x^\prime,D,\lambda,N_{\mathrm{e}}(x)]$ in the case where
$N_{\mathrm{e}}(x)$ follows a Gaussian distribution in $x$.  The
Gaussian distribution is a convenient function for describing a lens
that is localized to some region of characteristic size $a$ and whose
refractive strength declines quickly outside of this region.  In this
section we develop the refractive properties of the Gaussian lens in
the limit of geometrical optics.  We will show that these properties
can be parameterized in terms of a single dimensionless quantity that
is constructed from the physical parameters $\lambda$, $a$, $D$, and
$N_{\mathrm{e}}(x)$.

The column density profile takes the form
\begin{equation}
N_{\mathrm{e}}(x) = N_0 \exp\left[-(x/a)^2\right].
\label{eqn:nex}
\end{equation}
Following the discussion in \S\ref{sec:geo_optics}, the phase advance
due to such a lens is
\begin{equation}
\phi(x) = \lambda r_{\mathrm{e}} N_0 \exp\left[-(x/a)^2\right].
\label{eqn:phgauss}
\end{equation}
The refraction angle, equation~(\ref{eqn:ref_ang}), becomes
\begin{equation}
\theta_{\mathrm{r}}(x)
 = - (\lambda^2 r_{\mathrm{e}} N_0/ \pi a^2) x \exp\left[-(x/a)^2\right]
\label{eqn:thrgauss}
\end{equation}
and 
the ray path, equation~(\ref{eqn:ray_geom_skew}), is
\begin{equation}
x^\prime
 = x\left\{1 + (\lambda^2 r_{\mathrm{e}} N_0 D/ \pi a^2) \exp\left[-(x/a)^2\right] \right\}
 - \theta_{\mathrm{i}} D.
\label{eqn:raypath}
\end{equation}

Considerable simplification takes place if the following dimensionless
variables are defined:
\begin{mathletters}
\begin{eqnarray}
u                   & \equiv & x/a;         \\
u^\prime            & \equiv & x^\prime/a;  \\
\theta_\ell & \equiv & a/D;         \\
\beta_{\mathrm{s}}               & \equiv & \theta_{\mathrm{i}} / \theta_{\ell}; \\
\gamma              & = & u^\prime + \beta_{\mathrm{s}}.
\end{eqnarray}
\end{mathletters}
The angle $\theta_{\ell}$ is roughly the angular width of the
lens as seen by the observer, while $\beta_{\mathrm{s}}$ is the incidence angle of
the background rays upon the lens screen in units of
$\theta_{\ell}$.

In the above equations the physical parameters $\lambda$, $a$, $D$,
and~$N_0$ occur consistently in the combination $\alpha$,
\begin{eqnarray}
\alpha & \equiv & \lambda^2 r_{\mathrm{e}} N_0 D/ \pi a^2, \nonumber \\
       & = & \left(\frac{\sqrt{\lambda D}}{a}\right)^2\frac{1}{\pi}\lambda r_{\mathrm{e}}N_0.
\label{eqn:phys_alpha}
\end{eqnarray}
We have written $\alpha$ in this second form to emphasize the
essential physics.  The phase advance through the lens is $\lambda
r_{\mathrm{e}}N_0/\pi$.  The Fresnel scale is $\sqrt{\lambda D}$.
Thus, the properties of the lens are determined by the square of the
ratio of the Fresnel scale to the lens size and the phase advance
through the lens.  The larger the parameter $\alpha$, the greater are
the observable effects due to the lens.  Hence a weak lens, $\lambda
r_{\mathrm{e}}N_0/\pi \ll 1$, can produce large observable effects if
the Fresnel scale is sufficiently larger than the lens size.
Similarly, a strong lens, $\lambda r_{\mathrm{e}}N_0/\pi \ll 1$, need
not produce large observable effects if the Fresnel scale is small
relative to the lens size.  Numerically,
\begin{equation}
\alpha 
 = 3.6 \Biggl( \frac{\lambda}{1\,\mathrm{cm}} \Biggr)^2\Biggl( \frac{N_0}{{1\,\mathrm{cm}^{-3}\,{\mathrm{pc}}}} \Biggr)\Biggl( \frac{D}{1\,\mathrm{kpc}} \Biggr)\Biggl( \frac{a}{1\,\mathrm{AU}} \Biggr)^{-2}.
\label{eqn:alpha}
\end{equation}

Upon substitution of the above dimensionless variables, the following
expressions are derived for the refractive properties of a Gaussian
lens:

\begin{description}
\item[refraction~angle]
\begin{equation}
\theta_{\mathrm{r}}(u)/\theta_{\ell} = - \alpha u \exp(-u^2);
\label{eqn:refrac_ang}
\end{equation}
\item[ray~path]
\begin{equation}
u[1+\alpha \exp(-u^2)] - \gamma = 0;
\label{eqn:u_geom}
\end{equation}
\item[gain~factor]
\begin{equation}
G_k = [1 + (1-2u_k^2)\alpha\exp(-u_k^2)]^{-1};
\label{eqn:gain_fac}
\end{equation}
\item[total~intensity]
\begin{equation}
I(u^\prime,\alpha) 
 = \sum_{k=1}^{n} \int_{-\infty}^{+\infty} B(\beta_{\mathrm{s}})G_k(u^\prime,\alpha,\beta_{\mathrm{s}})\,d\beta_{\mathrm{s}}.
\label{eqn:total_int}
\end{equation}
\end{description}


\section{Characteristic Light Curve Produced by a Gaussian Lens}
\label{sec:lt_curve}

When transverse motion between source, lens, and observer is assumed,
e.g., motion of the observer along the $u^\prime$ axis, $I (u^\prime,
\alpha)$ translates into a light curve $I(u^\prime, \alpha; t, v)$
since $u^\prime(t) = u^\prime(t=0) + vt$ where $t$ is time and $v$ is the
relative transverse velocity.  We will show that interpretation of
radio light curves in terms of plasma lenses that can be approximated
as Gaussian in profile can lead to inferences regarding the physical
properties of the lens.

\subsection{General Description}\label{sec:gen_descrip}

We will present numerical results in following sections.  Here we will
first develop a physical understanding for the basic characteristics
of refraction by a Gaussian plasma lens, Fig.~\ref{fig:gauss_lens}.
Plane waves from an infinitely distant source are incident on the lens
screen.  The plane waves are indicated by straight dotted lines in the
figure, and the lens is represented by a plot showing the electron
column density as a function of coordinate $u$ along the lens plane.
This representation is purely schematic, as the lens is assumed to
have a negligible but uniform width along the line of sight.  Upon
emergence, the constant phase surfaces are distorted into contours
that mimic the function $N_{\mathrm{e}}(u)$, as represented by the
inverse Gaussian shaped dotted lines.  The lines are inverse Gaussian
shaped because the phase velocity is greatest through the center of
the lens, where $N_{\mathrm{e}}$ reaches its maximum value of $N_0$.

In the limit of geometrical optics, the rays of energy flux travel
perpendicular to the constant phase surfaces.  The direction of travel
of the rays is indicated schematically by the small arrows in the
figure.  Extending this concept, we have drawn the path of
approximately 100 rays as they travel perpendicular to the constant
phase surface after emergence from the lens.  The ray path is shown
for a total distance $D$ along the lens axis.  An observer located
close to the lens plane but far from the lens axis (e.g., the upper
left and upper right regions of the ray trace) sees an unchanged
source: There is no change in the observed flux density (no change in
the number density of the rays) or in the source's position (the
direction of arrival of the rays).

Somewhat farther from the lens and slightly off-axis, in the
``Focusing Regions'' marked in Fig.~\ref{fig:gauss_lens}, the rays
begin to converge and their number density increases.  An observer
located in this area of the ray trace would see a source of enhanced
brightness somewhat displaced from its ``true'' position, as evident
from the increased number density and skewness of the rays,
respectively.  At the same distance from the lens, but closer to the
lens axis, the rays are spread apart by the lens.  An observer in this
region would see a source of decreased brightness due to the lower
number density of rays.  The source would be offset from its true
position by an amount that increases with the observer's distance from
the lens axis.  Directly on the lens axis, the number density of rays
reaches a minimum, but the source appears in its correct position.

At a larger distance from the lens plane, the skewed rays that make up
the focusing regions cross one another.  Inside this region where the
rays have crossed, more than one bundle of rays would be incident on
an observer.  Each bundle would appear to come from a different
direction.  The observer therefore sees more than one image of the
background source.  In the case of a Gaussian lens, the greatest
number of images seen is three.  Since each ray bundle arrives from a
different part of the lens, the spreading of the rays in each bundle,
and therefore the brightness of the particular image represented by
each bundle, may differ.

Outside the focusing region, only a single image is seen.  At large
distances from the lens axis, the image has its nominal brightness, as
evident from the uniform density and straight path of the rays in this
region.  Close to the lens axis the image has a reduced brightness as
evident from the spreading of rays in this region.

Separating the single image regions from the multiple image regions
are the caustic surfaces.  On these surfaces, an observer sees
multiple images merge into a single image, and the local number
density of rays from that image grows without bound in the limit of
geometrical optics.  In the case of a Gaussian lens, two of the three
images merge on the caustic surface.  In reality, the brightness of
the image is limited by diffraction effects which are not taken into
account here.

Now assume that an observer, due to relative transverse motion of the
lens, travels at a constant (large) distance~$D$ from the lens plane.
The transverse coordinate along the observer's path is given by
$u^\prime$, which is a function of time, $t$.  As the observer
encounters different ray densities, the brightness of the background
source (and its apparent position) will change with time.  At the
bottom of Fig.~\ref{fig:gauss_lens} is a light curve constructed by
plotting the number density of rays encountered as the observer
travels across the bottom of the ray trace.  The intensity $I$ is
normalized to unity in the absence of the lens.  In the multiple
imaging regions, the observed brightness is taken as the sum of the
number densities of the ray bundles that reach the observer, in
accordance with equation~(\ref{eqn:total_I}).

When the observer crosses a caustic surface, the number density of
rays tends towards infinity (for geometrical optics), and therefore
the observed brightness of the source becomes very large.  In our
refraction simulations, this increase in intensity is represented by
tall spikes in the light curve.  The height of the spikes is generally
not properly represented in refraction simulations, but qualitatively
their existence is correct.

\subsection{Numerical Results}\label{sec:num_result}

\subsubsection{Point Source}\label{sec:pt_src}

We start by considering the refraction of a background point source by
a Gaussian lens.  In this case $\beta_{\mathrm{s}} \equiv 0$, and we
take $I_0$ to be the intensity of the background source in the absence
of lensing.  The observed intensity, equation~(\ref{eqn:total_int}),
reduces to
\begin{equation}
I(u^\prime,\alpha) = I_0 \sum_{k=1}^n G_k(u^\prime,\alpha).
\label{eqn:intens_sum}
\end{equation}
For a given value of $\alpha$, the function $I(u^\prime,\alpha)$ is
determined in practice by numerically computing the root(s) $u_k$ to
equation~(\ref{eqn:u_geom}) (which is a transcendental function) for a
particular $u^\prime$, then computing the gain factor(s) $G_k$ of
equation~(\ref{eqn:gain_fac}) and performing the sum in
equation~(\ref{eqn:intens_sum}).  This is done for a range of
$u^\prime$ to construct the light curve.

As an initial example, consider a Gaussian plasma lens with peak
electron column density $N_0 = 0.1$~pc~cm${}^{-3}$ and width $a =
2$~\hbox{AU}.  The observer is located a distance $D = 1$~kpc away and
is observing at a wavelength $\lambda = 20$~cm.  For this particular
combination of physical parameters, $\alpha = 36$,
equation~(\ref{eqn:alpha}); other combinations can also produce
similar values of $\alpha$.
Equations~(\ref{eqn:refrac_ang})--(\ref{eqn:total_int}) have been
solved over the range $-20 \leq u^\prime \leq 20$ for this value of
$\alpha$, and $\sum G_k$ for each $u^\prime$ is plotted in
Fig.~\ref{fig:lcurve}.  The plot represents the enhancement (or
decrease) in the brightness of the background point source over its
value in the absence of a lens.  The features in the light curve
follow the general discussion in the previous section.  On the lens
axis, $u^\prime = 0$, there is one image and the intensity is at a
minimum, with a gain factor of $G_1 = 0.027$ in accordance with
$1/(1+\alpha)$ as predicted by equation~(\ref{eqn:gain_fac}).

Other values of $\alpha$ will give different light curves.  Larger
$\alpha$ gives rise to greater transverse displacement of the rays.
This can occur by (1)~increasing the distance $D$ between observer and
lens; (2)~increasing the maximum column density $N_0$ or decreasing
the characteristic size $a$ so that the transverse gradient of the
column density, and therefore the refraction angle, becomes greater;
or (3)~observing at a longer wavelength, where the index of refraction
$n_{\mathrm{r}}$ becomes smaller (greater deviation from 1).

Light curves are shown for various values of $\alpha$ in
Fig.~\ref{fig:ltcurve_alphas}.  Consider first the cases of weak
refraction, $\alpha = 0.1$ and $\alpha = 1$.  The transverse
displacement of the rays is small and no ray crossing occurs.
Modulations in the light curve are due solely to focusing and
defocusing of the ray bundles since there are no caustics.  When
$\alpha$ is increased to 5, the onset of ray crossing is evident from
the existence of caustic spikes.  However, the observer is close to
the distance $D$ where ray crossing just begins to occur.  This is
evident from the close proximity between the inner and outer pairs of
caustics, compared to the width of the defocusing region near the lens
axis (cf.\ Fig.~\ref{fig:gauss_lens}).  We will say more in
\S\ref{sec:caustics} on the distances between caustics and what that
tells us about the lens.  Moderate refraction is shown by the example
$\alpha = 10$.  The pairs of caustics have moved farther apart and the
minimum in the light curve on the lens axis becomes deeper: $1/11$ of
its nominal value.  The light curve for $\alpha = 40$ exhibits strong
refraction.  The outer pair of caustics forms very far from the lens
axis, near the limits of the plot, while the inner pair is not as
strongly affected.  This fact will be exploited in
\S\ref{sec:caustics} to estimate the size of the lens based on
observed light curves.  On-axis, the source is only $1/41$ of its
nominal brightness.

\subsubsection{Extended Source}\label{sec:extend_src}

For an extended source the effect of the lens is determined not only
by $\alpha$, but also by the relative angular extent of the source
compared to the angular extent ($\sim a/D$) of the lens.  Since we
will generally invoke a Gaussian brightness distribution across the
source, the parameter $\beta_{\mathrm{s}}$ is taken as the ratio of the FWHM
angular width of the source to the observed angular FWHM of the lens.

The numerical simulations are shown in Fig.~\ref{fig:ltcurve_betas}.
Using an $\alpha=25$ lens we show the light curve from a point
background source, $\beta_{\mathrm{s}} = 0$, as well as those for $\beta_{\mathrm{s}} = 0.1$,
0.25, 0.5, 1.0, and~5.0.  It is evident that an extended source
smoothes out the caustic spikes and the transition into the defocusing
region near the lens axis.  However, the brightness minimum at the
center of the lens remains substantial until the source is larger than
the lens.  For instance, when $\beta_{\mathrm{s}} = 1$ the on-axis gain is 0.039,
only 2\% larger than the gain $1/(1+\alpha)$ predicted by a point
source.  When the source is much bigger than the lens the result of
refraction is to produce a ripple in the light curve rather than
strong modulations.  Rays from the outer edges of the source are
refracted into the defocusing region so the minimum intensity is not
as deep as for a point source.

Note that the height of the ripple and the true effects on the caustic
spikes of an extended source can not be quantitatively determined
without a simulation that also includes diffraction effects.  We have
performed limited diffraction simulations and can say that the
refractive light curves are correct in a qualitative sense.

\section{The Caustics}\label{sec:caustics}

The caustics are a powerful probe of the physical properties of the
lens.  In our refractive simulations the important observable is the
distance between the pairs of caustic spikes.  Diffractive fringes
form in the vicinity of the caustics due to interference between the
multiple images.  Though it is beyond the scope of this paper, the
fringes can serve as a useful probe of both the lens and the
background source (e.g., \cite{wc87}).

As evident from Fig.~\ref{fig:ltcurve_alphas}, the distance between
the inner two caustics measured in units of the characteristic size
$a$ of the lens is a relatively weak function of $\alpha$, while the
distance between the outer two caustics depends more strongly on
$\alpha$.  We have solved for the locations where the caustics
intersect the $u^\prime$-axis, as a function of $\alpha$. We define
the dimensionless quantities $\Delta u^\prime_{\mathrm{i}}$ and
$\Delta u^\prime_{\mathrm{o}}$ as the distances between, respectively,
the inner and outer pairs of caustics, in dimensionless units of
$u^\prime$.  These quantities are plotted as functions of $\alpha$ in
Fig.~\ref{fig:caus_sep}, where we have also plotted $(\Delta
u^\prime_{\mathrm{o}} - \Delta u^\prime_{\mathrm{i}})/2$, the distance
between the outer and inner caustics on each side of the lens axis.
It is evident from the figure that no caustics form if
$\log_{10}\alpha < \log_{10}\alpha_{\mathrm{min}} \simeq 0.35$
($\alpha_{\mathrm{min}} \simeq 2.25$).  In this case the observer is
closer to the lens than the point at which ray crossing occurs. For
$\alpha = \alpha_{\mathrm{min}}$, caustics have formed but the inner
and outer caustics are merged together.  As $\alpha$ increases, the
outer pairs spread apart while $\Delta u^\prime_{\mathrm{i}}$ remains
relatively fixed.

Although these separations are derived from solutions to
transcendental equations, it is possible to find an analytic
expression for $\Delta u^\prime_{\mathrm{o}}$ as a function of
$\alpha$, the derivation of which we present in the next section.  For
$\Delta u^\prime_{\mathrm{i}}$ we have performed a linear
least-squares fit in the log-log domain to find an approximate
functional dependence on $\alpha$.  The separations are 
\begin{eqnarray}
\Delta u^\prime_{\mathrm{i}} & \approx & 4.30 \alpha^{0.055}; \\
\label{eqn:deltaui}
\Delta u^\prime_{\mathrm{o}} & = & \sqrt{2}\left(1 + \alpha e^{-1/2}\right).
\label{eqn:deltauo}
\end{eqnarray}

The dimensionless quantity $\Delta u^\prime_{\mathrm{i}}$ is $\Delta
x^\prime_{\mathrm{i}} / a$, where $\Delta x^\prime_{\mathrm{i}}$ is
the distance between the inner two caustics.  The characteristic size
$a$ of the lens can therefore be estimated as $a \simeq \Delta
x^\prime_{\mathrm{i}}/4.3$.  Additionally, the value of $\alpha$ can
be estimated by measuring the relative separations between the inner
and outer pair of caustics:
\begin{equation}
\alpha 
 \simeq 1.7\left(3\frac{\Delta x^\prime_{\mathrm{o}}}{\Delta x^\prime_{\mathrm{i}}} - 1\right)^{1.06},
\label{eqn:est_alpha}
\end{equation}
where $\Delta x^\prime_{\mathrm{o}}$ is the distance between the outer
pair of caustics.

We emphasize that the distance between the inner pair of caustics is
the \emph{only reliable estimate for the size of the lens}.  Using the
value $\Delta x^\prime_{\mathrm{o}}$ determined through observations
of the total duration of the lensing event to judge the size of the
lens is not appropriate, cf.\ Fig.~\ref{fig:caus_sep}.  It can deviate
by a factor of order $\alpha$ from the true characteristic size $a$ of
the lens.  Since physically plausible values of $\alpha$ can exceed
$10^4$, a characteristic lens size derived from the total duration of
the lensing event will overestimate the true size by a comparable
factor.

\section{Angular Displacement}\label{sec:ang_dis_sec}

In addition to changes in the observed brightness of the lensed
source, the plasma lens also causes variations in the apparent angular
position of the source.  Consider an observer at a point $u^\prime$,
who observes a ray that, as a solution to equation~(\ref{eqn:u_geom}),
comes from the point $u_k$ on the lens plane.  In the small angle
approximation, the angular displacement of the source from its
``true'' (unlensed) position is $(u^\prime - u_k)
\theta_{\ell}$, cf.\ equation~(\ref{eqn:ray_geom}).  The value
of $(u^\prime - u_k)$ depends on $u$ and the parameters $\alpha$ and
$\beta_{\mathrm{s}}$.  For simplicity, we will restrict our discussion to the case
of a point source, for which $\beta_{\mathrm{s}} = 0$.

Figure~\ref{fig:ang_dis} shows the general behavior of the angular
displacement of a background source as viewed through a plasma lens.
While this figure was computed for one particular value of $\alpha$,
the qualitative form of the angular displacement is independent of
$\alpha$.  The figure shows the function $(u^\prime - u_k)$, for valid
solutions of equation~(\ref{eqn:u_geom}) as a function of $u^\prime$.
The lens axis intersects the observer's axis at $u^\prime = 0$.  The
vertical axis, both in contour plot form and in a
pseudo--three-dimensional form, is the relative intensity of the image
coming from a particular value of $(u^\prime - u_k)$.  At values of
$u^\prime$ far from the lens axis, e.g., $u^\prime = 5$, $(u^\prime -
u_k) = 0$, and there is no angular displacement (Image~A in the
figure).  Closer to the lens axis, the outer caustic is formed, e.g.,
near $u^\prime = 3.1$.  In Fig.~\ref{fig:ang_dis}, the caustic is
evident by the sudden formation of a second, bright image at a new
value of $(u^\prime - u_k)$.  Image~A also remains; that is,
$(u^\prime - u_k)$ as a function of $u^\prime$ becomes multi-valued.
The new bright image is angularly displaced from the ``true'' position
of the background source as evident from the non-zero value of
$(u^\prime - u_k)$.

Even closer towards the lens axis, the image from the caustic breaks
up into two separate images (B and~C in the figure), and their
intensities diminish.  Image~B moves towards the original image
(Image~A) as $u^\prime$ decreases, i.e., as one moves closer to the
lens axis.  The intensity of Image~C diminishes quickly.

Somewhat closer to the lens axis, the inner caustic forms, e.g.,
$u^\prime \approx 2.6$, where Images~A and~B merge into a single very
bright image.  Then, Images~A and~B disappear altogether, leaving only
the weak Image~\hbox{C}.  This image moves towards zero angular
displacement as the observer nears the lens axis, $u^\prime =0$.  On
the axis, Image~C has zero angular displacement and it reaches its
minimum flux density, $1/(1+\alpha)$ of the nominal source intensity.

The magnitude of the angular displacement can be estimated from
equation~(\ref{eqn:ray_geom}).  The angular displacement in radians
for an image from a point $u_k$ on the lens plane reaching the point
$u^\prime$ on the observer plane is
\begin{equation}
\delta \theta = (u^\prime - u_k)\,\theta_{\ell},
\label{eqn:displace}
\end{equation}
where we have assumed that the lens axis passes through $u = u^\prime
= 0$.  For a Gaussian lens, the ray path description,
equation~(\ref{eqn:u_geom}), gives $u^\prime$ in terms of the
solution(s) $u_k$,
\begin{equation}
(u^\prime - u_k) = u_k \alpha e^{-u_k^2}.
\label{eqn:diff_u}
\end{equation}
The maximum value of the difference $(u^\prime - u_k)$ occurs for $u_k
= 1/\sqrt{2}$, which is a solution to equation~(\ref{eqn:u_geom}) when
the observer is located at the point $u^\prime = (\alpha e^{-1/2} +
1)/\sqrt{2}$.  The total separation between the two outer caustics is
twice this value, from which equation~(22) is derived.
This expression is true for all $\alpha$.

In Fig.~\ref{fig:ang_dis} it is evident that the maximum angular
displacement occurs as the observer crosses the outer caustic.
Images~B and~C are merged into one image there.  Since the image is
coming from the point $u_k = 1/\sqrt{2}$ and the observer is at the
point $u^\prime = (\alpha e^{-1/2} + 1)/\sqrt{2}$, the angular
displacement of the merged image is
\begin{eqnarray}
\delta \theta & = & (u^\prime - u_k) (a/D), \nonumber \\
 & = & \frac{1}{\sqrt{2}} \alpha \theta_{\ell} e^{-1/2}.
\label{eqn:delta_theta}
\end{eqnarray}

\section{Application to Extreme Scattering Events}\label{sec:ese}

We have presented a quantitative analysis of the refractive properties
of a specific optical system: a one-dimensional interstellar plasma
lens. The specific case of a Gaussian distribution of electron density
transverse to the line of sight has been considered.  We do not
believe that such ideal systems exist in reality, but the results of
our analysis provide a starting point for a semi-quantitative analysis
of extreme scattering events towards distant radio sources.

Fiedler et al.~(1994a) summarized the variety of light curves observed
during ESEs.  The scattering events are pronounced at lower
frequencies ($\approx 1$~GHz), but usually weakly detected or not
detected at higher frequencies ($\sim 10$~GHz), presumably due to a
combination of weaker ISM scattering at shorter wavelengths and
intrinsic noise in the radio light curves.  One exception to this
general rule is the first detected scattering event, towards the
quasar 0954$+$658, which occurred in 1981. At 8.1~GHz, the most
prominent features in the light curve are four strong spikes,
reminiscent of caustic spikes due to a point-like source being
refracted by a strong lens.

The lack of caustic spikes in most of the identified ESEs, however,
implies that the typical plasma lenses that produce ESEs are fairly
weak (small $\alpha$), so that multiple imaging does not occur; the
lenses have much smaller angular diameters than the background sources
(large $\beta_{\mathrm{s}}$), so that caustic surfaces are less pronounced; or a
combination of both effects.  Fey, Clegg, \& Fiedler~(1996) present
multi-epoch, multi-wavelength images of sources in which an ESE has
been identified at times when the sources were not undergoing an
\hbox{ESE}.  They find the sources to be compact, with typical FWHM
diameters of approximately 1~mas at wavelengths between 3.5 and~18~cm.
Figure~\ref{fig:ltcurve_betas} indicates that, if the lack of caustic
spikes is because the lenses have smaller angular diameters, the
typical lens diameter would be $\theta_{\ell} \lesssim 0.1$~mas
(0.1~AU at 1~kpc).  We therefore favor the weak-lens model as the most
plausible explanation for the general lack of caustic spikes in ESEs.

The best-studied ESE is that for the extragalactic source 1741$-$038
which occurred in 1992.  High quality light curves were obtained at
2.25 and 8.1~GHz during the event and refraction effects were detected
in both.  Radio polarization data obtained during the event and 1.5~yr
afterwards have been used to constrain the magnetoionic structure of
the lensing medium (Clegg et al.~1996).  VLBI observations of
1741$-$038, also obtained outside the ESE, have been used to determine
the unlensed angular structure of this source.  Fey et al.~(1996)
used these data in conjunction with the Fielder et al.~(1994a) model
to infer that the angular diameter of the lens and the source must be
comparable ($\approx 0.5$~mas).

We have chosen the ESEs towards 0954$+$658 and 1741$-$038 for detailed
analysis with respect to our Gaussian lens model.  This choice was
made due to the apparent formation of caustics during the 0954$+$658
event and the quality of the data obtained during the 1741$-$038
event.  In the following discussion, we will obtain semi-quantitative
estimates of lens parameters and indicate how additional data would
be useful in improving our estimates.

\subsection{0954$+$658}\label{sec:0954+658}

The 2.25 and 8.1~GHz light curves obtained for 0954+658 are shown in
Fig.~\ref{fig:0954_lt2}.  The ESE, which occurred in 1981, is shown in
Fig.~\ref{fig:0954_lt} on an expanded scale.  The 2.25~GHz ESE is
marked by a flat-bottomed minimum, at which the flux density of the
source is one-half its nominal (unlensed) value, surrounded by two
peaks, which we identify below as the inner two caustics.  The 8.1~GHz
ESE is marked by a 2~Jy peak and three 1~Jy peaks bracketing an
approximately 1.5~month minimum in the source's flux density.  The
minimum is less pronounced at 8.1~GHz, as the source's flux density
during the event decreases to 70\% or so of its nominal value.  Below, we
identify these four, approximately 1~Jy peaks as the caustics.

The flat-bottomed minimum in the 2.25~GHz ESE indicates that, at this
frequency, the intrinsic angular diameter of 0954$+$658 is
substantially smaller than the angular diameter of the plasma lens,
cf.\ Fig.~\ref{fig:ltcurve_betas}.  However, the fact that the flux
density during the ESE only decreases to one-half its nominal value
can be reconciled with our lens model only if the light curve is the
sum of a lensed and an unlensed component.  If we take the amplitude
of the light curve during the minimum to be the flux density of the
unlensed component, the light curve can be represented by a
two-component source model in which a component of flux density
0.35~Jy is lensed by the intervening medium and a second component of
flux density 0.3~Jy is unaffected by the lens; the angular diameter of
the lensed component also has to be sufficiently small enough to
produce the flat-bottomed minimum.

The unlensed component could be either a compact component of
0954$+$658, such as a jet component, whose line of sight was
unaffected by the lens, or emission on larger scales that is
completely resolved by VLBI arrays, but is unresolved to the much
smaller Green Bank Interferometer used to obtain the light curve.
Both scenarios are consistent with the observed structure of
0954$+$658.

Gabuzda et al.~(1992, 1994) conducted 5~GHz observations between 1987
and 1989, and found the source to consist of a compact core ($<
0.5$~mas) with a jet extending approximately 5~mas to the northwest.
Within the jet they identified multiple, compact ($< 1$~mas) knots.
With the possible exception of the knot K0 (\cite{gmcwr94}), however,
none of the knots they identified would have been present at the time
of the \hbox{ESE}.  The proper motion of the knots is approximately
0.4~mas~yr${}^{-1}$, large enough that most of the knots were not
ejected until after the ESE (epoch 1981.1), unless the knots have
undergone substantial acceleration since their ejection.  Pearson \&
Readhead~(1988) included this source in a finding survey conducted in
1978.  Though they did not image it, they found it to be compact and
poorly modelled as a single gaussian component.  These VLBI
observations bracket the time of the ESE and found indications of
compact, yet complex structure.  We therefore conclude that it is
plausible that, at the time of the ESE, the source consisted of
multiple compact components, whose typical angular scales are
milliarcseconds.

On intermediate to larger angular scales, tens of milliarcseconds to
arcseconds, there is also emission from this source.  Gabuzda et
al.'s~(1992, 1994) VLBI observations also included simultaneous VLA
observations.  They found that approximately 15\% ($\approx 0.15$~Jy)
of the flux was resolved out by the VLBI observations.  Also, using
lower resolution VLA observations Kollgaard et al.~(1992) find a jet
extending approximately 4\arcsec\ to the south of the VLA core; the
jet's flux density is approximately 0.22~Jy.  This extended emission
would lie entirely within the synthesized beam of the Green Bank
Interferometer, but would be unaffected by the passage of a Gaussian
lens in front of the source.  In summary, this source displays
emission on a range of angular scales, from submilliarcsecond to
arcseconds.  A combination of small-scale components, whose line of
sight the lens did not intersect, and emission on larger scales could
provide the 0.35~Jy of unlensed flux.

To obtain quantitative estimates of the lens' parameters, we proceed
as follows.  We obtain approximate values for $\alpha$ and
$\beta_{\mathrm{s}}$ by appeal to the light curves.  We refine our
estimates for $\alpha$ and $\beta_{\mathrm{s}}$ by producing simulated
light curves with slightly different values of $\alpha$ and
$\beta_{\mathrm{s}}$ and comparing visually the simulated and actual
light curves.  Due to the simple model we are considering here, we
judged a more exhaustive search of the $(\alpha, \beta_{\mathrm{s}})$
parameter space not to be worthwhile.

Comparing the shape of the inner caustics and minimum of the 2.25~GHz
ESE to Fig.~\ref{fig:ltcurve_betas}, we estimate $\beta_{\mathrm{s}}$ to be $0.1 <
\beta_{\mathrm{s}} < 1$.  From the separation of the inner and outer caustics in
the 8.1~GHz ESE and equation~(\ref{eqn:est_alpha}), we estimate
$\alpha \approx 10$ at this frequency.  Since $\alpha \propto
\lambda^2$, at 2.25~GHz we have $\alpha \approx 130$.  

Varying the parameters and comparing simulated light curves to the
actual 2.25~GHz light curve, we find reasonable agreement for $\alpha
= 160$ and $\beta_{\mathrm{s}} = 0.4$.  In Fig.~\ref{fig:0954_lt} we
have superposed the simulated light curve for this set of parameters
on the 2.25~GHz light curve of 0954$+$658.  In doing so, we have
scaled the $u^\prime$ axis arbitrarily since it depends on the unknown
relative transverse velocities of the observer and lens.  The
agreement with the ESE minimum is good.  The simulation underestimates
the amplitude of the inner pair of caustics.  The outer caustics, for
$\alpha = 160$, are far outside the time range of the observations
plotted in Fig.~\ref{fig:0954_lt}.  We have compared our simulation to
the full time range of the observations of 0954$+$658,
Fig.~\ref{fig:0954_lt2}.  We do not see any evidence for the existence
of the outer caustics, but the predicted amplitude of the outer
caustics is significantly smaller than the predicted amplitude for the
inner caustics.  The outer caustics are probably lost in the noise in
the light curve.  The ESE occurs during a general decrease in the
source's flux density, probably caused by intrinsic variations.  This
general decrease in the source's flux density explains the lack of
agreement between the amplitude of the actual light curve and the
model light curve after the \hbox{ESE}.

To compare our model with the 8.1~GHz light curve, we scale the light
curve fit to the 2.25~GHz ESE to 8.1~GHz.  We believe such a simple
scaling is justified within the context of our 1-D simulations of a
Gaussian lens passing in front of a single compact component of
0954$+$658.  Below we discuss briefly the consequences of relaxing the
assumptions in our simple model.

We assume that the angular size of the lensed component of 0954$+$658
scales as $\lambda$, as is generally appropriate for flat-spectrum
extragalactic radio sources.  Though the refractive index within the
lens does change with wavelength, the physical size of the lens should
not.  Thus, the relative angular extent of the lensed component 
is $(2.25/8.1) = 0.28$ times smaller at 8.1~GHz as compared to
2.25~GHz or $\beta_{\mathrm{s}} = 0.11$ at 8.1~GHz.  Our estimated value of
$\alpha$ at 2.25~GHz, $\alpha = 160$, corresponds to $\alpha = 12$ at
8.1~GHz.  The resulting model light curve is superposed on the actual
light curve in Fig.~\ref{fig:0954_lt}.  The best match for the flux
densities of lensed and constant components at 8.1~GHz is 0.15
and~0.45~Jy, respectively.

We identify tentatively the lensed component of 0954$+$658 as a jet
component, not the core.  The difference in flux density at the two
frequencies for the lensed component, 0.35~Jy at 2.25~GHz to 0.15~Jy
at 8.1~GHz, indicates that the lensed component has a steep spectrum,
$S \propto \nu^{-0.7}$.  The emission from the source at~5 and~8~GHz
is dominated by compact components (\cite{gcrw92}, 1994; \cite{gc96};
unpublished VLBI observations), generally with flat spectra (viz.\
Fig.~\ref{fig:0954_lt2}).  However, one of the two knots in the jet,
knot K3 (\cite{gmcwr94}; \cite{gc96}) has a similar spectral index
with a 5~GHz flux density of 0.2~Jy.  We re-iterate that, because of
proper motions, none of the knots seen at~5~GHz by Gabuzda et
al.~(1994) or at~8~GHz by Gabuzda \& Cawthorne~(1996) would have been
present at the time of the \hbox{ESE}.  Nevertheless, we believe the
existence of such a jet component at a later date suggests that such a
component could have been present during the \hbox{ESE}.  Clearly VLBI
observations during and after an ESE would be quite useful in
assessing which component(s) was lensed.

Qualitatively, Fig.~\ref{fig:0954_lt} shows that our simulation
reproduces the four caustic spikes in the 8.1~GHz light curve.
However, the time scale for the scattering event derived from our
model differs substantially from the observed data.  The model
predicts a duration of approximately 0.4 yr, while the observed
duration is about half that.  Equation~(\ref{eqn:deltaui}) and
Fig.~\ref{fig:caus_sep} show that the distance (and therefore time
scale) between the inner pair of caustics is a very weak function of
$\alpha$ in our model, and, therefore, a very weak function of
observing wavelength.  In essence, the time scale between the inner
caustics at 2.25 and~8.1~GHz should be nearly identical.  In the
observational data, that is not the case.

The most likely explanations for the discrepancy between our model and
the data are due to the limitations of a one-dimensional lensing
simulation.  The wavelength scaling of the timescale of the lensing
event derived in the 1-D model does not account for the possibility
that the mid-point of the lens may not cross the background source.
Instead, only a chord of the lens, rather than the full diameter, may
pass in front of the background source.  Combined with complex
wavelength-dependent structure in the background source, meaning that
the predominant contribution to the flux density may shift in position
with wavelength, the net effect will be a complicated wavelength
scaling of the time scale for the lensing event.

We use our estimates of $\alpha$ and $\beta_{\mathrm{s}}$ to constrain
the physical properties of the lens.  Since the better comparison is
obtained at 2.25~GHz, we will use the parameter values found at that
frequency: $\alpha = 160$, $\beta_{\mathrm{s}}=0.4$.  From
equation~(\ref{eqn:alpha}), the maximum electron column density
through the lens is
\begin{equation}
N_0
 = 0.28\,\mathrm{pc\,cm}^{-3}\,\alpha\lambda^{-2}D\theta_{\mathrm{s}}^2\beta_{\mathrm{s}}^{-2}.
\label{eqn:N0}
\end{equation}
Here $\theta_{\mathrm{s}}$ is the angular size of the background
source in units of milliarcseconds, and we have used the fact that
$\beta_{\mathrm{s}} \equiv \theta_{\mathrm{s}} / \theta_{\ell}$ from
equation~(13).

The unknown quantities in equation~(\ref{eqn:N0}) are the angular
diameter of the background source, $\theta_{\mathrm{s}}$, and the
distance to the lens, $D$.  The angular diameter is
$\theta_{\mathrm{s}} \lesssim 1$~mas (\cite{gcrw92}, 1994;
\cite{gc96}; unpublished VLBI data).  Toward 0954$+$658 is Galactic
Loop~III, which Fiedler et al.~(1994b) identify as responsible for the
\hbox{ESE}.  Its distance is estimated to be 0.15~kpc.  More
generally, the scale height of the Galaxy's free electron layer is
approximately 1~kpc (\cite{tc93}).  At the Galactic latitude of
0954$+$658 ($b = 43\fdg1$), the maximum distance to the lens would be
roughly 1.5~kpc.

With these estimates for $\theta_{\mathrm{s}}$ and $D$, we compute a
nominal column density through the lens of $N_0 \simeq
0.24$~pc~cm${}^{-3}$ corresponding to a distance of 0.15~kpc and an
upper limit of $N_0 \lesssim 2.4$~pc~cm${}^{-3}$ corresponding to a
distance of 1.5~kpc.  Our estimates for $\theta_{\mathrm{s}}$ and
$\beta_{\mathrm{s}}$ also provide constraints on the size of the lens.
We estimate $\theta_{\ell} \approx 2.5$~mas, corresponding to $a
\approx 0.38$~AU (3.8~AU) for a lens at a distance of 0.15~kpc
(1.5~kpc).  The free electron density within the lens is
$n_{\mathrm{e}} \approx N_0/a$ or $n_{\mathrm{e}} \sim
10^5$~cm${}^{-3}$; in our model, the electron column density estimate
is distance independent.  The mass of the lens is $M_\ell \sim
m_{\mathrm{p}}n_{\mathrm{e}}a^3$, where $m_{\mathrm{p}}$ is the mass
of the proton; for a lens at 0.15~kpc (1.5~kpc), $M_\ell \sim 1.3
\times 10^{20}\,\mathrm{g} \sim 6.5 \times
10^{-14}\,\mathrm{M}_{\sun}$ ($1.3 \times 10^{23}\,\mathrm{g} \sim 6.5
\times 10^{-11}\,\mathrm{M}_{\sun}$).

The angular displacement of the position of the source during the ESE
can be estimated from equation~(\ref{eqn:delta_theta}), $\delta\theta
\simeq 250$~mas at 2.25~GHz.  Phase-referenced VLBI observations can
obtain absolute position information.  Angular displacements of this
magnitude should be detectable easily, even if the phase-referencing
source is a few degrees away (\cite{bc95}).  In the case of
0954$+$658, the radio reference frame source 0951$+$693 (M81) is
roughly 4\arcdeg\ away. Of course, if the source is not lensed
completely, relative positions between components can be determined
with high accuracy without recourse to phase-referenced observations.
We stress that the maximum angular displacement, and the optimal time
for such observations, occurs during passage through the outer
caustic.

We close this section speculatively.  First, we use the linear lens
size and the time scale of the ESE to estimate the velocity of the
lens.  From equation~(\ref{eqn:deltaui}), $v \approx 4.3a/\Delta t$.
The time scale between the inner caustics of the ESE is $\Delta t
\approx 0.25$~yr.  Thus, the velocity of the lens is 30~km~s${}^{-1}$
(300~km~s${}^{-1}$) for a lens at 0.15~kpc (1.5~kpc).  Second, the
lens responsible for the 0954$+$654 ESE is required to be a factor of
a few larger than the source; $\beta_{\mathrm{s}} = 0.4$ implies the
angular diameter of the lens is a factor of 2.5 greater than the
angular diameter of the lensed source component.  However, the lens'
diameter is also comparable with the size of the jet imaged by Gabuzda
et al.~(1992, 1994).  If the lensed component was indeed a jet
component, the lens would have had to pass in front of the jet
component without affecting the flat-spectrum core.  We therefore
suggest that the lens was anisotropic, with the major axis
corresponding to the derived lens diameter.  The minor axis would have
to be smaller than the typical knot separation in the jet, probably
less than 1~mas, suggesting an axial ratio of at least 2.5:1.  We simply
note that anisotropic structures in the interstellar plasma are not
unexpected (e.g., \cite{h84}, 1986).  Finally, the smooth, single
minimum in the 2.25~GHz light curve suggests that the lens encountered
only a single component in the jet, i.e., moved northeast-southwest
(or vice versa), roughly perpendicularly across the jet axis.

\subsection{1741$-$038}\label{sec:1741-038}

The complete light curve for 1741$-$038 is shown in
Fig.~\ref{fig:1741-038} with the ESE, which occurred in 1992, shown in
Fig.~\ref{fig:1741_lt} on an expanded scale.  At both frequencies the
ESE displays a rounded minimum, with the 2.25~GHz ESE causing an
approximately 50\% decrease in the source's flux density and the
8.1~GHz ESE causing an approximately 30\% decrease in the source's
flux density.  The lack of a flat bottom and the apparent lack of
caustic spikes at either frequency suggest that 1741$-$038 was only
weakly lensed, i.e., no multiple images were formed and $\alpha <
\alpha_{\mathrm{min}}$.

Our analysis for 1741$-$038 proceeded in the same fashion as that for
0954$+$658.  Our best model comparison to the 2.25~GHz light curve
gives the following parameter values: $\alpha = 2$, $\beta_{\mathrm{s}} = 1.0$,
with a single-component (lensed) source model of flux density 2~Jy.
The model is superposed on the observational data in
Fig.~\ref{fig:1741_lt} and replicates the minimum of the 2.25 GHz
event very well.  It overestimates the amplitude of the maxima
surrounding the minimum, particularly the one preceding the decrease
in the source's flux density, but the amplitudes of these maxima are
not too much larger than the level of (presumably) intrinsic
fluctuations in the light curve.

Scaling the value of $\alpha$ to 8.1~GHz, we find $\alpha = 0.14$.
Scaling the source size by $\lambda$ gives $\beta_{\mathrm{s}} \simeq
0.28$, but Fey et al.~(1996) show that the dominant component of
1741$-$038 is unresolved at this frequency.  We therefore assume that
$\beta_{\mathrm{s}} = 0$ at 8.1~GHz (in practice, there is very little
difference between $\beta_{\mathrm{s}} = 0.28$ and $\beta_{\mathrm{s}}
= 0$ light curves).  The single component model with $\alpha = 0.14$,
$\beta_{\mathrm{s}} = 0$, and lensed source flux density of 2.55~Jy is
shown in Fig.~\ref{fig:1741_lt}.  The observed depth of the light
curve is greater than that reproduced by the model and the observed
minimum occurs slightly before that predicted by the model.  These
deficiencies may indicate that the lens did not pass directly over the
source, crossing only at a grazing incidence, that the electron column
density within the lens is more complicated that the Gaussian form we
have assumed, or that the source contains unresolved substructure.

Following our analysis for 0954$+$658, we require the source size and
distance to the lens in order to constrain $N_0$,
equation~(\ref{eqn:N0}).  Fey et al.~(1996) show that the angular size
of 1741$-$038 is approximately 0.5~mas at 2.32~GHz.  Since
$\beta_{\mathrm{s}} = 1$, the lens must also be approximately 0.5~mas
in diameter.  A number of sources displaying ESEs, including
1741$-$038, are seen along the edges of radio Loop~I (\cite{fdjws94}).
Berkhuisjen~(1973) estimates a distance of 130 $\pm$ 75~pc to
Loop~\hbox{I}.  If we assume the ESE arises from a structure
associated with Loop~I, the resulting peak electron column density
through the lens is $N_0 \simeq 10^{-4}$~pc~cm${}^{-3}$.  If we do not
associate the ESE with Loop~I, we can place a limit on its distance
only by requiring that the lens be within one scale height of the free
electron layer of the Galaxy.  This distance limit is $D \lesssim
4.5$~kpc, for a peak column density of $N_0 \lesssim 3.6 \times
10^{-3}$~pc~cm${}^{-3}$.  Although a connection between Loop~I and the
ESEs is suggestive, the line of sight to 1741$-$038 (and other, nearby
sources which have undergone an ESE) passes through the inner Galaxy
and some other Galactic structure may be responsible for the ESEs.

The inferred size of the lens is 0.065~AU ($\lesssim 2.25$~AU) if the
lens is (is not) associated with Loop~\hbox{I}.  The density within
the lens is $n_{\mathrm{e}} \approx 300$~cm${}^{-3}$, for a
corresponding mass of $1.6 \times 10^{15}\,\mathrm{g} \sim 8 \times
10^{-19}\,\mathrm{M}_{\sun}$ ($6.8 \times 10^{19}\,\mathrm{g} \sim 3.4
\times 10^{-14}\,\mathrm{M}_{\sun}$) if the lens is (is not)
associated with Loop~\hbox{I}.  The time scale between the inner
caustics is 0.4~yr.  The corresponding velocity of the lens is
3~km~s${}^{-1}$ ($\lesssim 110\,\mathrm{km\,s}^{-1}$) for a lens
associated (not associated) with Loop~\hbox{I}.

These lens parameters are comparable to those derived for the ESE for
PSR~B1937$+$21 (\cite{cblbadd93}).  $N_0 = 6.5 \times
10^{-5}$~pc~cm${}^{-3}$ and $a \approx 0.07$~\hbox{AU}.  (They
actually fit for two lenses, we quote the total electron column
density through both lenses and the mean size of the two lenses.)

As we indicated in \S\ref{sec:ang_dis_sec}, angular displacement
occurs regardless of the value of $\alpha$.  Thus, even though caustic
surfaces were not formed during lensing of 1741$-$038, we can still
apply equation~(\ref{eqn:delta_theta}).  We find the maximum angular
displacement during the ESE to be approximately 0.4~mas at 2.25~GHz
and 0.03~mas at 8.1~GHz.  It is unlikely that these angular
displacements could be measured using phase-referenced observations.
Relative position shifts of this magnitude between lensed and unlensed
components might be detectable at 2.25~GHz, particularly if the source
structure could be determined soon after the ESE so that proper motion
of the components is negligible.

Clegg et al.~(1996) conducted polarization observations of 1741$-$038
with the VLA during and after the ESE in an effort to detect changes
in Faraday rotation as a result of the lens passing across the line of
sight.  No rotation measure change was detected to a level of
$\Delta\mathrm{RM} \lesssim 10$~rad~m$^{-2}$.  The rotation measure
change is given by
\begin{equation}
\Delta RM \simeq 0.81 \langle B_\parallel \rangle N_0,
\label{eqn:delrm}
\end {equation}
where $\langle B_\parallel \rangle$ is the mean component of the
magnetic field within the lens parallel to the line of sight in
microgauss. Our derived value of $N_0$ is consistent with the lack of
observed change in RM since the mean parallel magnetic field component
would have to be $\langle B_\parallel \rangle \gtrsim 120$~mG
(3.5~mG), if the lens is (is not) associated with Loop~I, in order for
a change in RM to have been detected.  These magnetic field levels are
$10^3$--$10^5$ times larger than those typical of the interstellar
medium.  If the lens is \emph{not} associated with Loop~I, Clegg et
al.~(1988) have shown that magnetic field levels $B \gtrsim 2$~mG can
be encountered as the result of strong interstellar shocks; however,
if the lens is associated with Loop~I, the magnetic field required to
produce an observable change in RM is still a factor of 60 larger than
what can plausibly be produced.

\section{Conclusions}\label{sec:conclude}

We have presented the geometric optics for refraction by an
interstellar plasma lens, with specific application to a lens with
Gaussian profile of free electron column density.  We have shown that
the one-dimensional refractive properties of a lens can be
characterized completely by two dimensionless parameters.  The first
parameter characterizes the refractive power of the lens and is
\begin{equation}
\alpha = 3.6 \left( \frac{\lambda}{1\,\mathrm{cm}} \right)^2
\left( \frac{N_0}{{1\,\mathrm{cm}^{-3}\,{\mathrm{pc}}}} \right)
\left( \frac{D}{1\,\mathrm{kpc}} \right)
\left( \frac{a}{1\,\mathrm{AU}} \right)^{-2}.
\eqnum{\ref{eqn:alpha}}
\end{equation}
where $\lambda$, $N_0$, $D$, and $a$ are the wavelength of
observation, maximum free electron column density, distance from lens
to observer, and size of the lens transverse to the line of sight,
respectively.  The second parameter characterizes the extent to which
the lens' effect is diminished by the intrinsic size of the source and is
\begin{equation}
\beta_{\mathrm{s}} = \theta_{\mathrm{s}} / \theta_{\ell},
\eqnum{13d}
\end{equation}
which is the relative angular size of the background source,
$\theta_{\mathrm{s}}$, compared to the angular size of the lens as
seen by the observer, $\theta_{\ell}$.

The effect of the plasma lens is to enhance or reduce the observed
brightness of the background source due to focusing or spreading of
ray bundles, and to change the apparent position of the background
source due to refraction.  The minimum brightness of the background
source occurs directly on the lens axis.  For a point source, $\beta_{\mathrm{s}} =
0$, the on-axis brightness will be $1/(1+\alpha)$ of its nominal
(unlensed) brightness.

On either side of the lens axis, at a distance determined by the value
of $\alpha$, the background source will have a maximum intensity due
to focusing of ray bundles.  If $\alpha \geq \alpha_{\mathrm{min}}
\simeq 2.25$, the focusing of ray bundles will become strong enough
for ray paths to cross, and multiple imaging of the background source
will occur.  For the case of a Gaussian lens profile, an observer
located within a region of multiple imaging can see as many as three
images of the background source, each with a different brightness and
appearing to come from slightly different directions.  The total
brightness of the source, obtained by adding the brightness of the
multiple images, will be greater than the nominal brightness of the
source.

Separating the multiple imaging regions from the single image regions
are caustic surfaces, on which two of the three multiple images blend
together and, in the limit of geometrical optics, the observed
brightness of the background source grows without bound.  An observer
passing from a single-image region through the multiple imaging
regions and back to a single-image region can use the observed
position of the caustics to determine the magnitude of
the parameter $\alpha$.  The separation between the \emph{inner}
caustics can also be used to constrain the diameter of the lens.

The primary effect of increasing the angular size of the background
source in relation to the angular size of the lens, i.e., increasing
$\beta_{\mathrm{s}}$, is to smooth out the light curve and reduce the observed
amplitudes of maximum and minimum intensity.

We have generated sample light curves from our model and compared them
to extreme scattering events observed towards the extragalactic
sources 0954$+$658, \S\ref{sec:0954+658}, and 1741$-$038,
\S\ref{sec:1741-038}.  These two sources were chosen because their
ESEs have the highest quality data available and, in the case of
1741$-$038, polarization observations also exist.  In general we find
reasonable agreement between the observed and modelled light curves at
2.25~GHz.  We have far more difficulty recreating the 8.1~GHz light
curves by utilizing only the appropriate wavelength scalings for
$\alpha$ and $\beta_{\mathrm{s}}$.  The discrepancies between the
modelled and observed light curves may result from some combination of
substructure within the lens, an anisotropic lens shape, a lens which
only grazes the source rather than passing completely over it, or
unresolved substructure within the extragalactic sources.

For 0954$+$658 we find it to be strongly lensed at 2.25~GHz.
Interpreting the light curve for this source is difficult because of
uncertainties about the source's structure at the time of the
\hbox{ESE}.  We favor an interpretation in which the lens passed over
a component in the source's jet.  The lens toward this source is
probably associated with radio Loop~III so that the lens was 0.15~kpc
distant.  The peak column density in the lens was 0.24~pc~cm${}^{-3}$,
its size was 0.38~AU, the electron density within the lens is
$10^5$~cm${}^{-3}$, and its mass was $6.5 \times
10^{-14}$~M${}_{\sun}$.  The lens also caused the source's position to
wander by as much as 250~mas at 2.25~GHz, an easily detectable amount
using phase-referenced VLBI observations or by measuring the relative
positions between source components.

For 1741$-$038, the ESE was caused by a weak lens.  The lens toward
this source is likely to be associated with radio Loop~I and at a
distance of 0.13~kpc.  The lens' peak column density is
$10^{-4}$~pc~cm${}^{-3}$, size is 0.065~AU, density is
300~cm${}^{-3}$, and mass is $10^{-18}$~M${}_{\sun}$.  From Clegg et
al.'s~(1996) upper limit on the change in Faraday rotation towards
1741$-$038 during the ESE, we place an upper limit of approximately
100~mG on the magnetic field within the lens.  The angular position
wander caused by this lens was only 0.4~mas at 2.25~GHz.  An angular
displacement of this magnitude would not have been detectable for
1741$-$038; many other sources that have undergone ESEs have more
complicated structures than 1741$-$038, however, and relative position
shifts of this magnitude between lensed and unlensed components might
be detectable in those sources.  These lens parameters are comparable,
within 50\%, to those derived for the ESE for PSR~B1937$+$21
(\cite{cblbadd93}).  $N_0 = 6.5 \times 10^{-5}$~pc~cm${}^{-3}$ and $a
\approx 0.07$~\hbox{AU}.  (They actually fit for two lenses, we quote
the total electron column density through both lenses and the mean
size of the two lenses.)

A key assumption of our plasma lens explanation for ESEs is that the
lenses are discrete objects.  The lens properties we have derived are
consistent with this assumption.  If we assume the plasma temperature
within the lenses to be $10^4$~K, the pressure within the lenses is
$nT \sim 3 \times 10^6$--$10^9$~K~cm${}^{-3}$.  These pressures are
well in excess of the average ISM pressure of roughly
4000~K~cm${}^{-3}$ (\cite{kh88}).  Such lenses would be either highly
transitory features or embedded in high-pressure environments.  Romani
et al.~(1987) suggest ionization fronts, cooling instabilities, or
both associated with old supernova remnants as sites of lenses.

We emphasize that the difficulties in interpreting the
light curve for future ESEs could be ameliorated considerably by VLBI
imaging of a source as soon as possible after, or indeed during, an
\hbox{ESE}.  Considerable angular displacements might also be detected
with such observations.

\acknowledgments
We thank J.~Cordes for guidance at the beginning of this project and
the referee, D.~Gabuzda, for comments which helped us clarify a number
of the points in this paper.  A portion of this work was performed
while TJWL held a National Research Council-NRL Research
Associateship.  Basic research in astronomy at the Naval Research
Laboratory is supported by the Office of Naval Research.

\clearpage

\begin{figure}
\caption[]{Example of ray path geometry through a plasma lens.
(\textit{a})~The ray from an infinitely distant point source is
incident normal to the lens plane, and strikes the lens plane at the
coordinate $x$.  Because of the non-uniform free electron column
density $N_{\mathrm{e}}(x)$, the lens refracts the ray through the
angle $\theta_{\mathrm{r}}$.  The ray travels the distance $D$ between
the lens plane and the observer's plane, and strikes the observer's
axis at the point $x^\prime$.  (\textit{b})~The same situation occurs,
except that the ray is incident with an oblique angle
$\theta_{\mathrm{i}}$ upon the lens plane, i.e., the infinitely
distant background point source is off-axis.}
\label{fig:lens_geom}
\end{figure}

\begin{figure}
\caption[]{Schematic diagram of refraction by a Gaussian plasma lens.
See \S\ref{sec:gen_descrip} for a complete description of this
figure.}
\label{fig:gauss_lens}
\end{figure}

\begin{figure}
\caption[]{Light curve of an infinitely distant point source refracted
by a lens with $\alpha = 36$.}
\label{fig:lcurve}
\end{figure}

\begin{figure}
\caption[]{Light curves produced by refraction through lenses with a
range of values of $\alpha$, as defined by
equation~(\ref{eqn:alpha}).}
\label{fig:ltcurve_alphas}
\end{figure}

\begin{figure}
\caption[]{Light curves produced by an $\alpha = 25$ lens, with
various relative angular sizes of background source and lens.  The
parameter $\beta_{\mathrm{s}}$ is the angular size of the (unlensed) background
source in units of the angular size of the lens, as seen by the
observer.}
\label{fig:ltcurve_betas}
\end{figure}

\begin{figure}
\caption[]{Separation between the outer caustics, $\Delta
u_{\mathrm{o}}^\prime$; between the inner caustics, $\Delta
u_{\mathrm{i}}^\prime$; and between the pairs of caustics on either
side of lens axis, $(\Delta u_{\mathrm{o}}^\prime - \Delta
u_{\mathrm{i}}^\prime)/2$, as functions of the parameter $\alpha$.}
\label{fig:caus_sep}
\end{figure}

\begin{figure}
\caption[]{Brightness and angular displacement of images formed by the
Gaussian plasma lens as functions of observer position $u^\prime$.
See \S\ref{sec:ang_dis_sec} for a complete description of this
figure.}
\label{fig:ang_dis}
\end{figure}

\begin{figure}
\caption[]{The complete light curve of the quasar 0954$+$658 during
the period of observation from 1979--1994.  The scattering event
occurs at approximately 1981.1.}
\label{fig:0954_lt2}
\end{figure}

\begin{figure}
\caption[]{The light curve of the quasar 0954$+$658 during an extreme
scattering event. Panel~(\textit{a}) shows the data obtained at
2.25~GHz, compared to our lens model (dashed line) with $\alpha = 160$
and $\beta_{\mathrm{s}} = 0.4$.  A two-component model for 0954$+$658
was used: a lensed component of flux density 0.35~Jy, and an unlensed
component of flux density 0.3~Jy.  Panel~(\textit{b}) is the 8.1~GHz
data compared to a two component model with $\alpha=12$,
$\beta_{\mathrm{s}} = 0.11$, lensed flux density 0.15~Jy, and unlensed
flux density 0.45~Jy.}
\label{fig:0954_lt}
\end{figure}

\begin{figure}
\caption[]{The complete light curve of the quasar 1741$-$038 during
the period of observation from 1983--1994.  The scattering event
occurs at approximately 1992.4.}
\label{fig:1741-038}
\end{figure}

\begin{figure}
\caption[]{The light curves of 1741$-$038 during an extreme scattering
event.  (\textit{a})~The dashed line is our model comparison with
$\alpha = 2$, $\beta_{\mathrm{s}} = 1$, and a flux density for
1741$-$038 of 2~Jy at 2.25~GHz.  (\textit{b})~The model uses $\alpha =
0.14$, $\beta_{\mathrm{s}} = 0$, and assumes an 8.1~GHz flux density
of 2.55~Jy.}
\label{fig:1741_lt}
\end{figure}

%
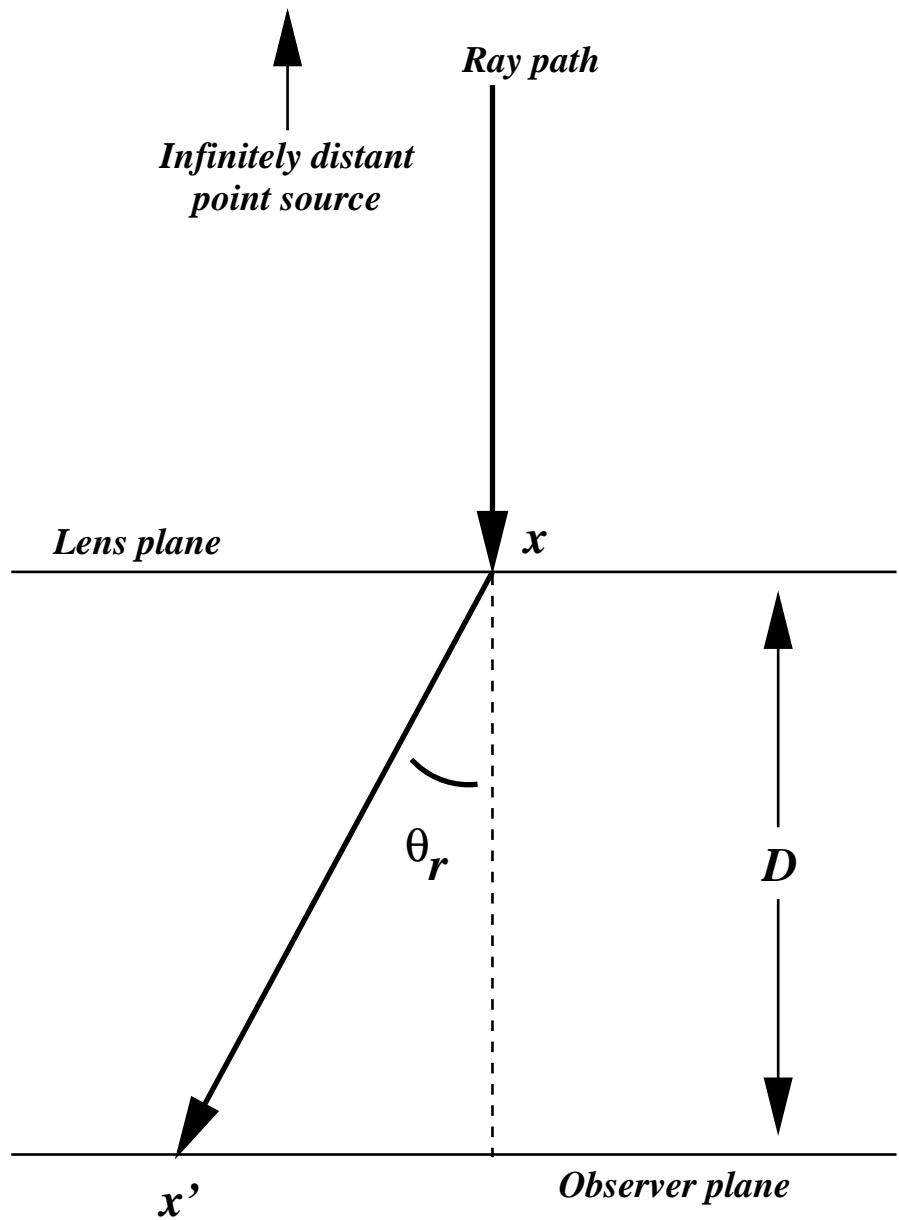
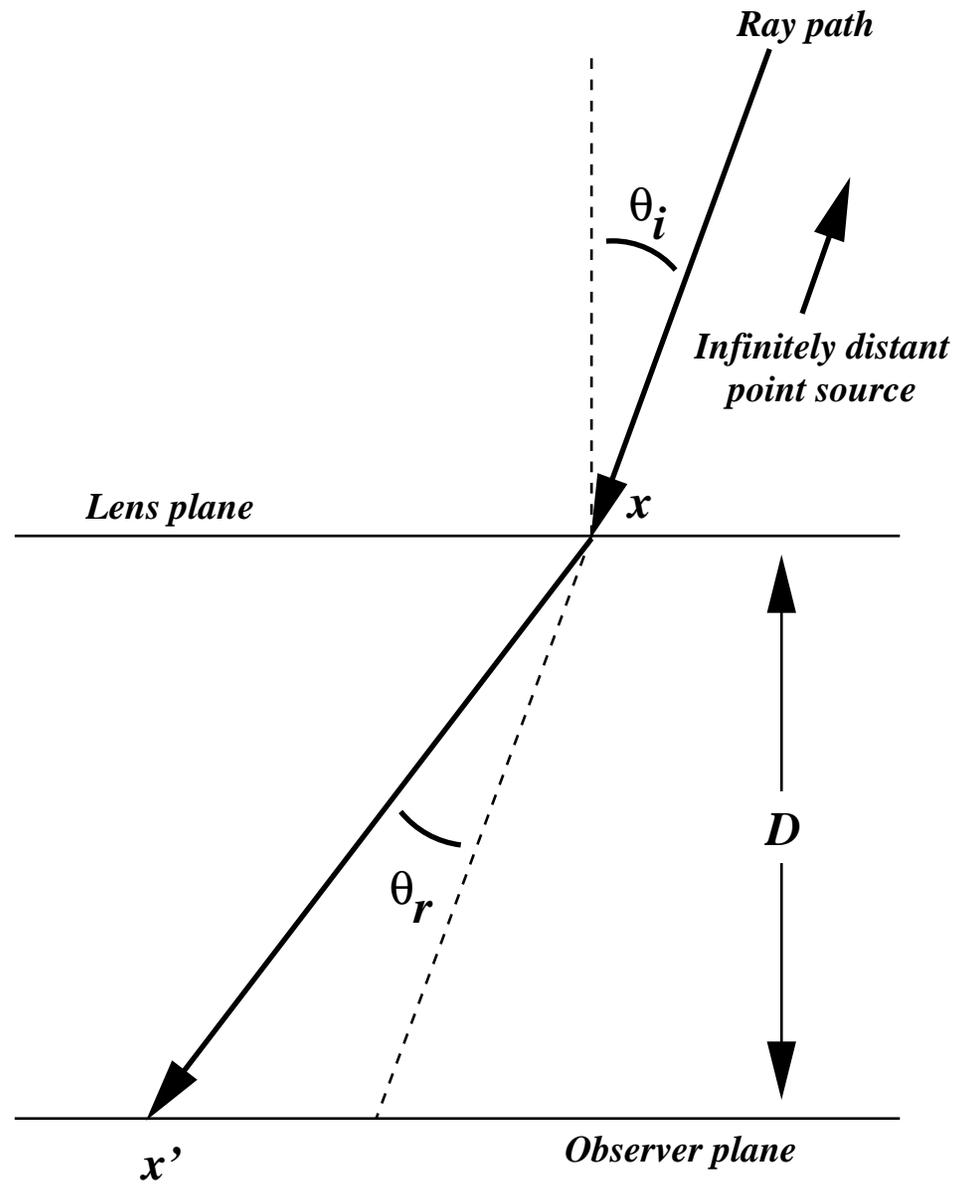
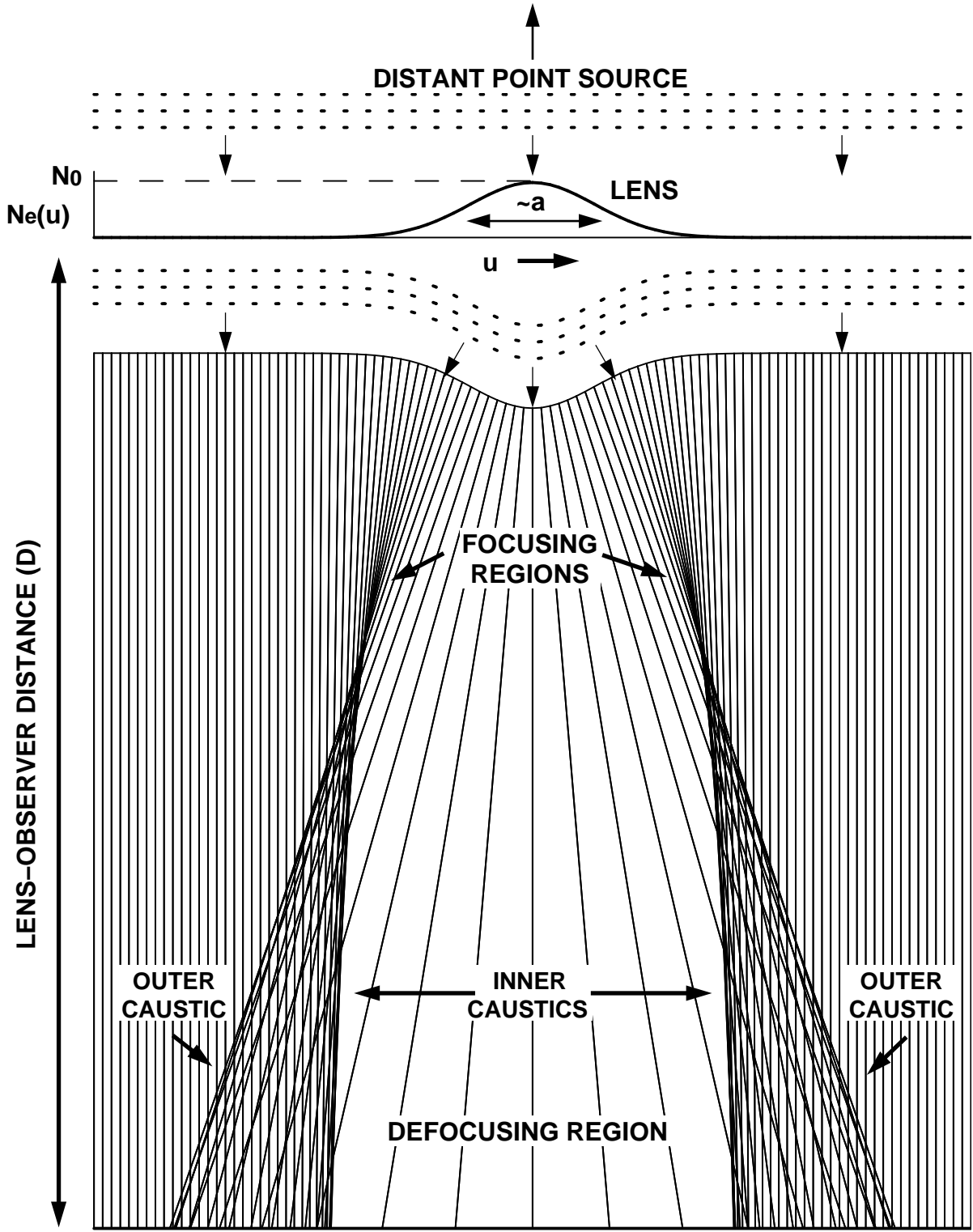
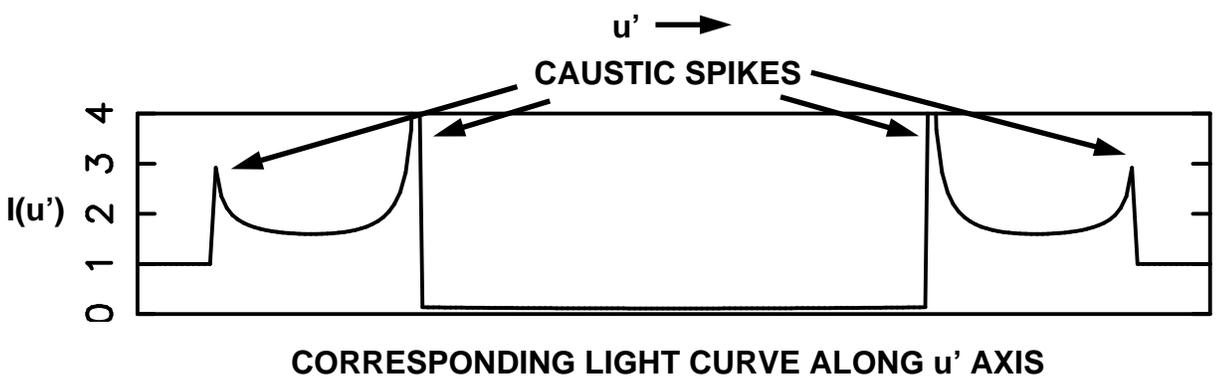
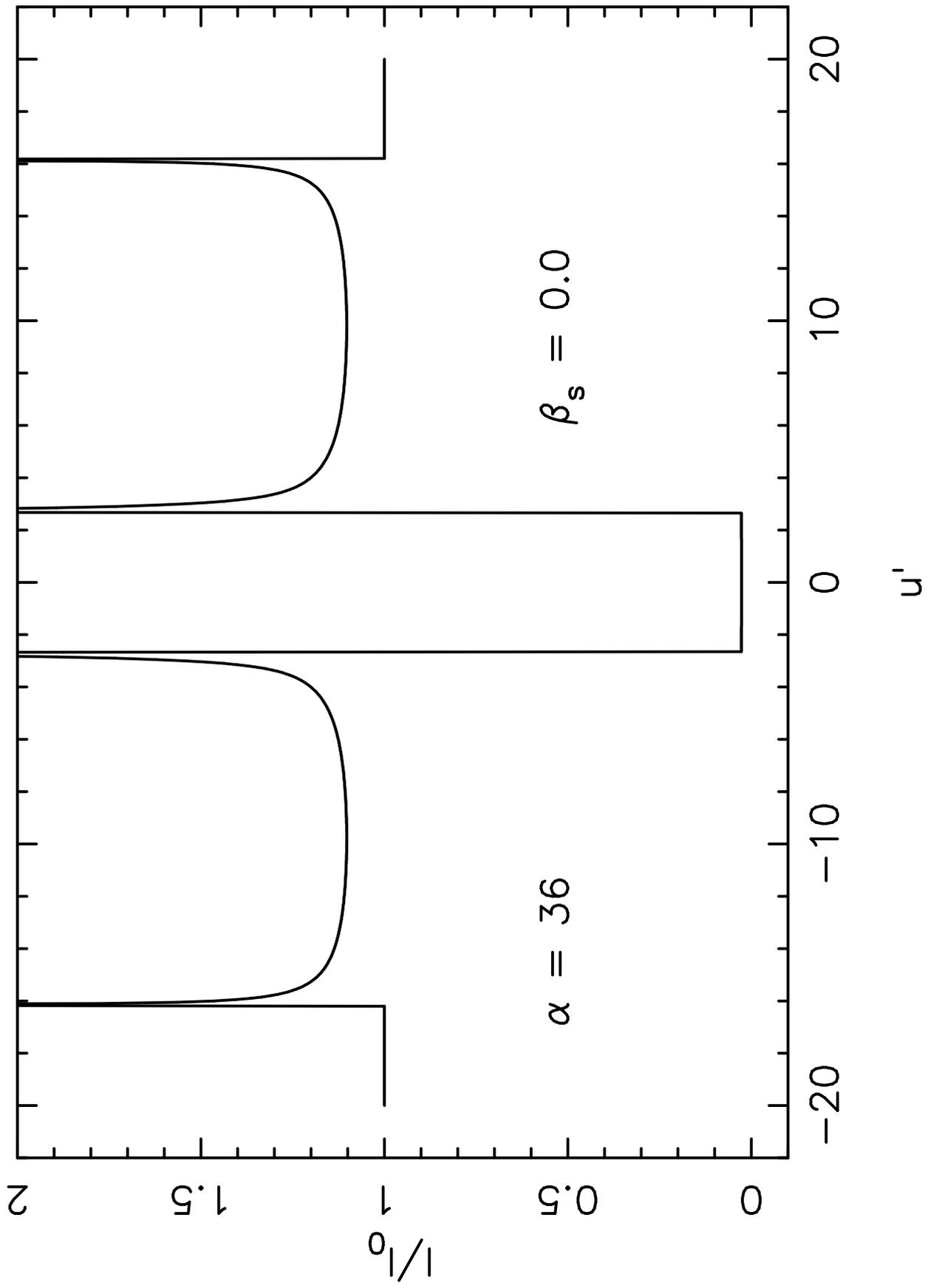
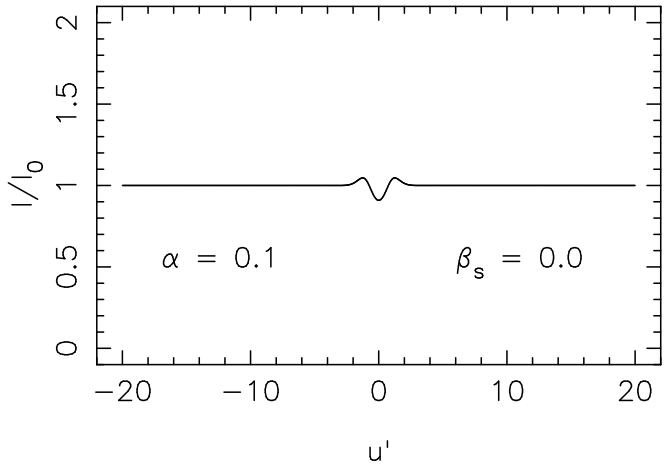
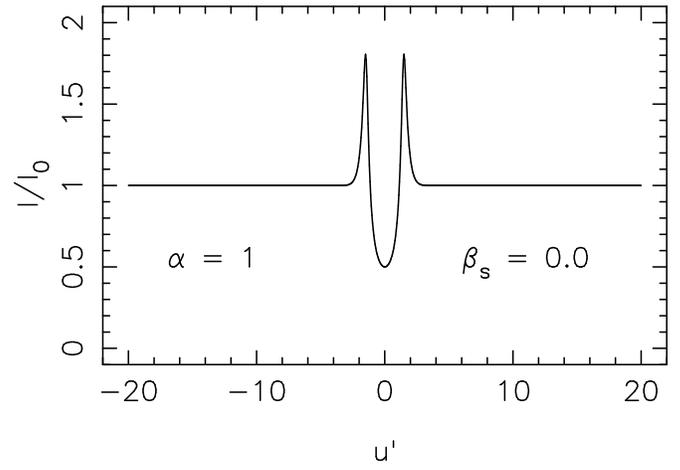
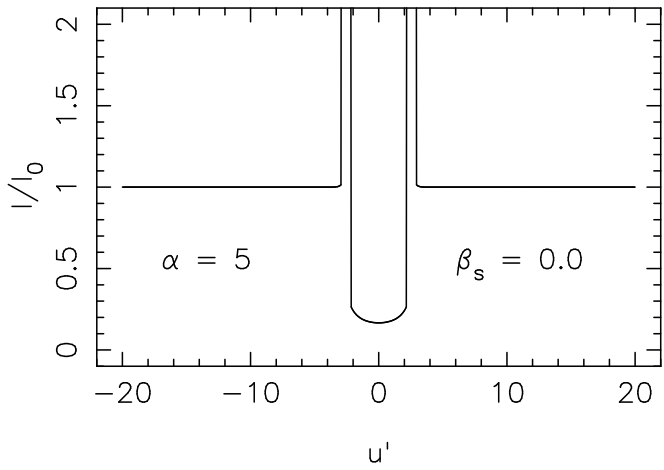
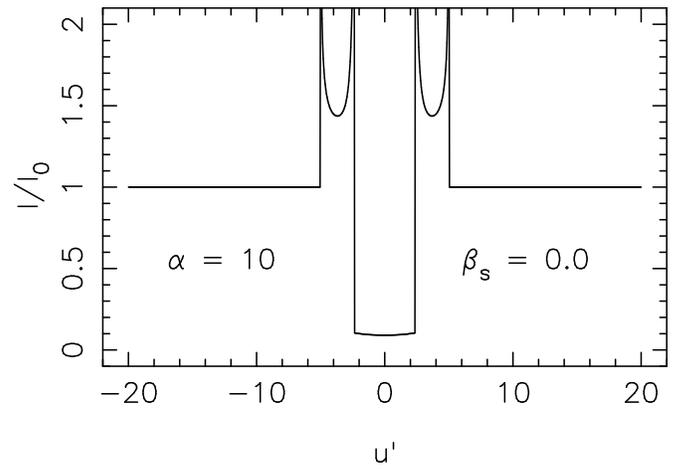
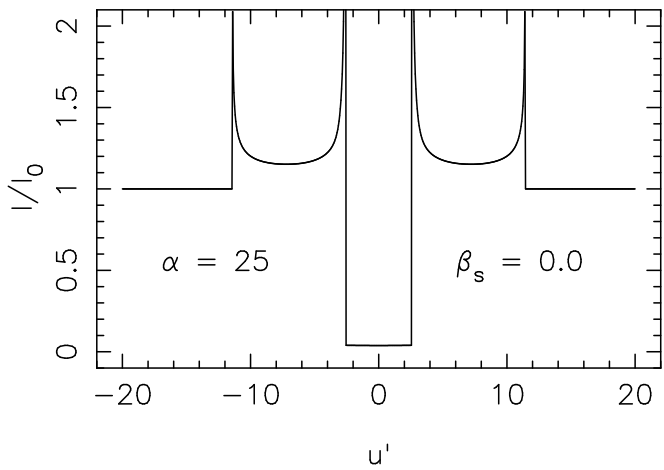
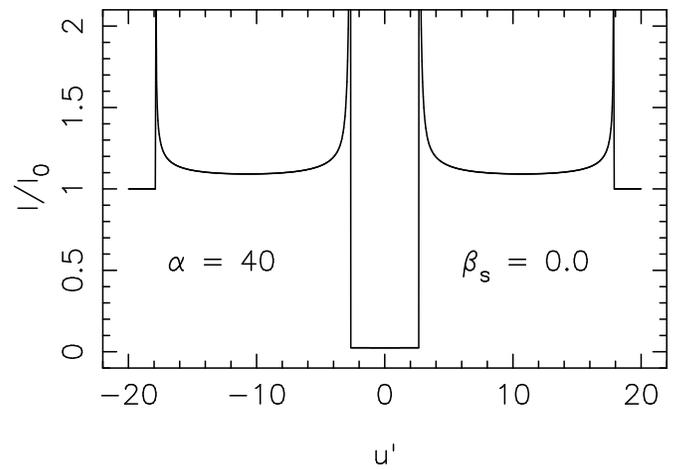
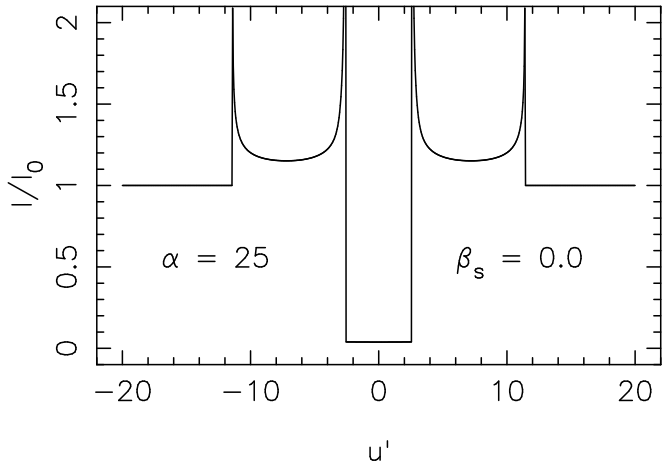
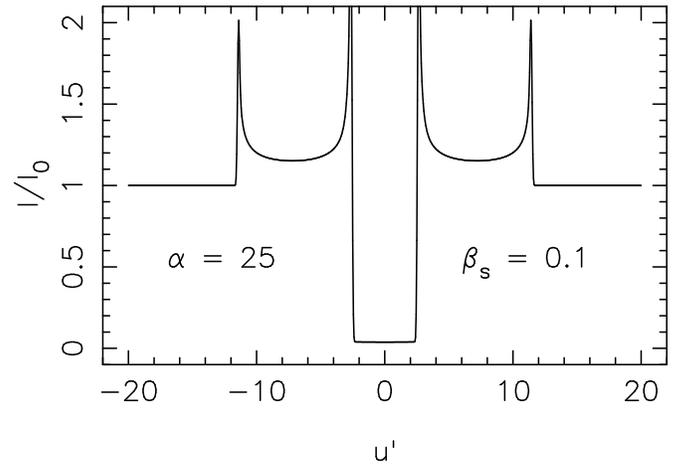
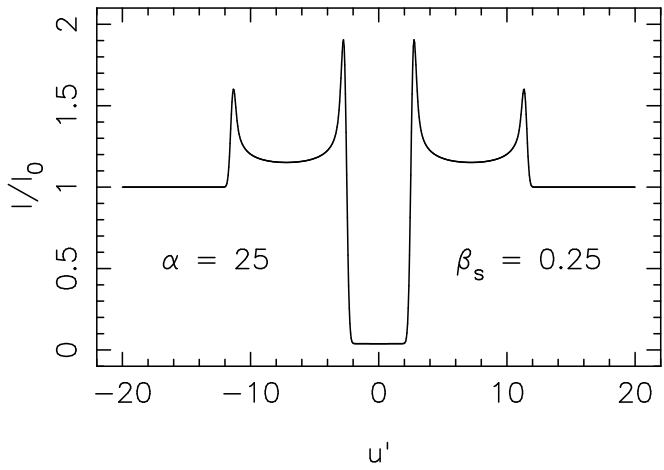
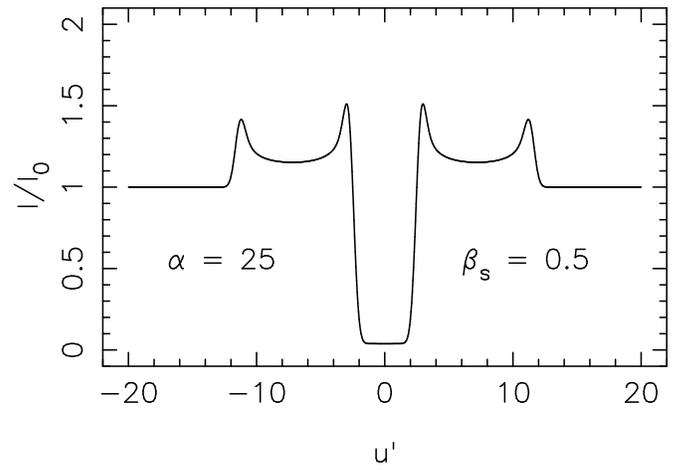
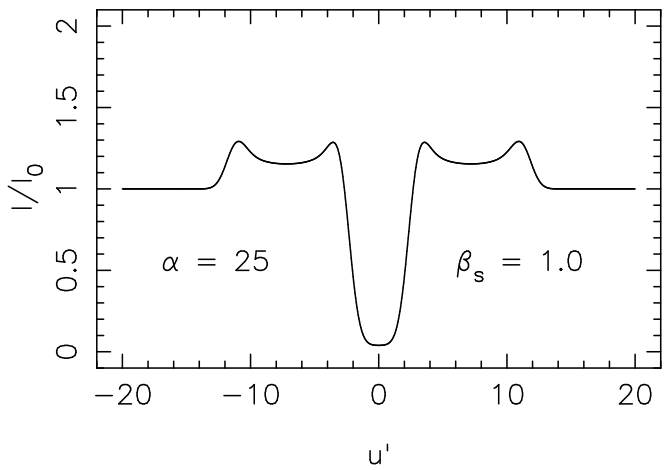
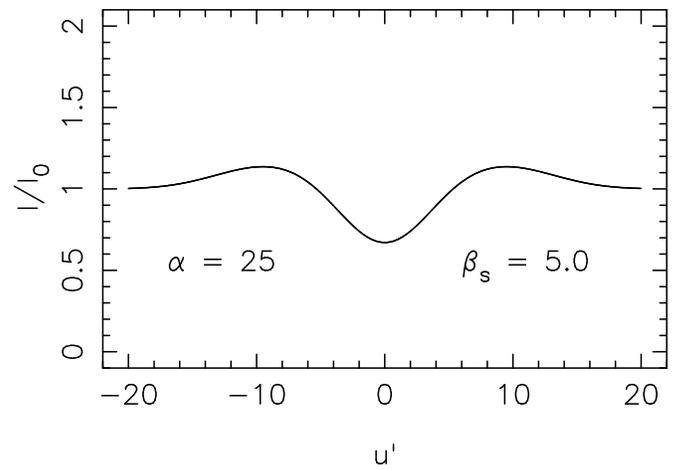
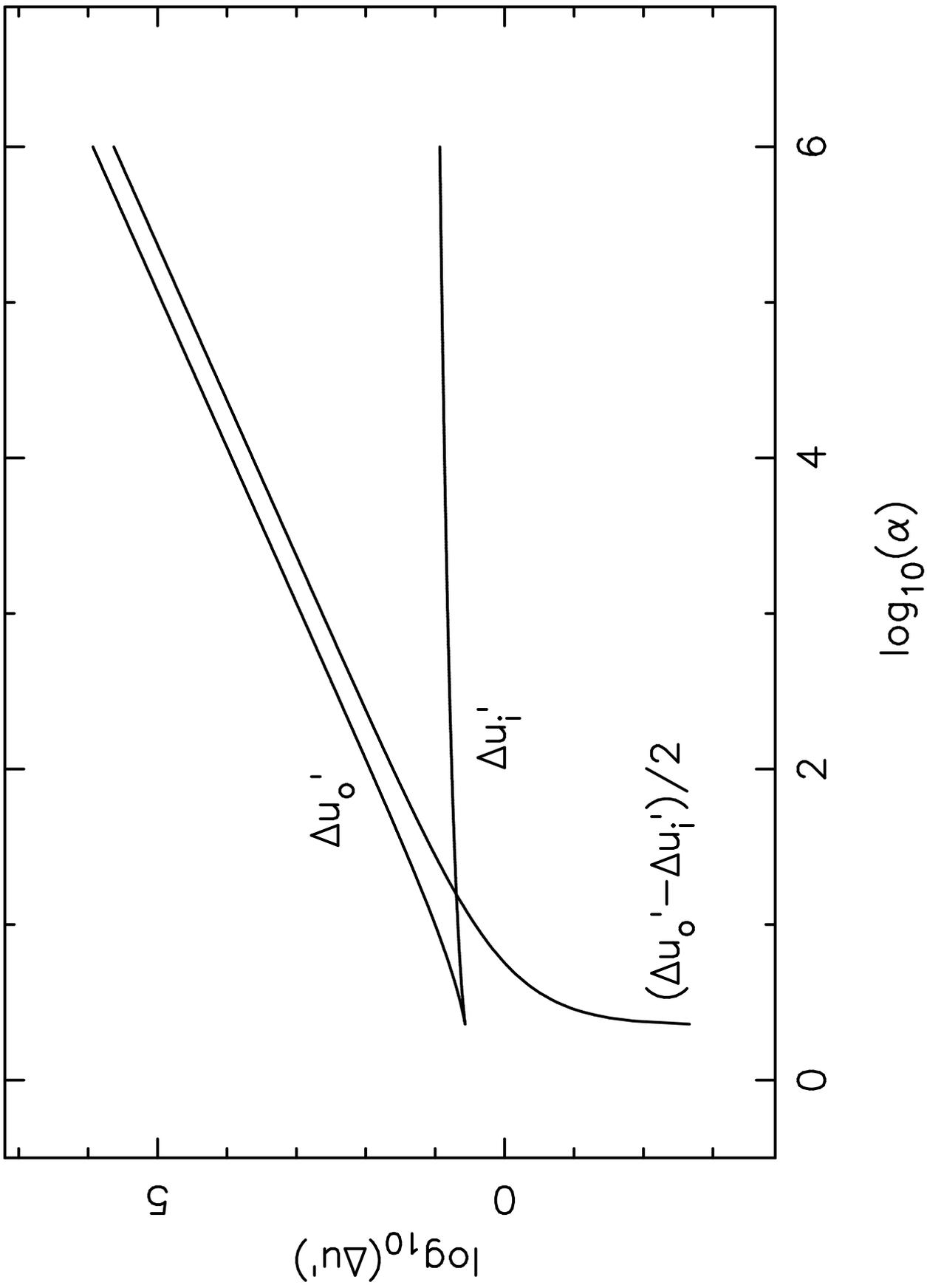
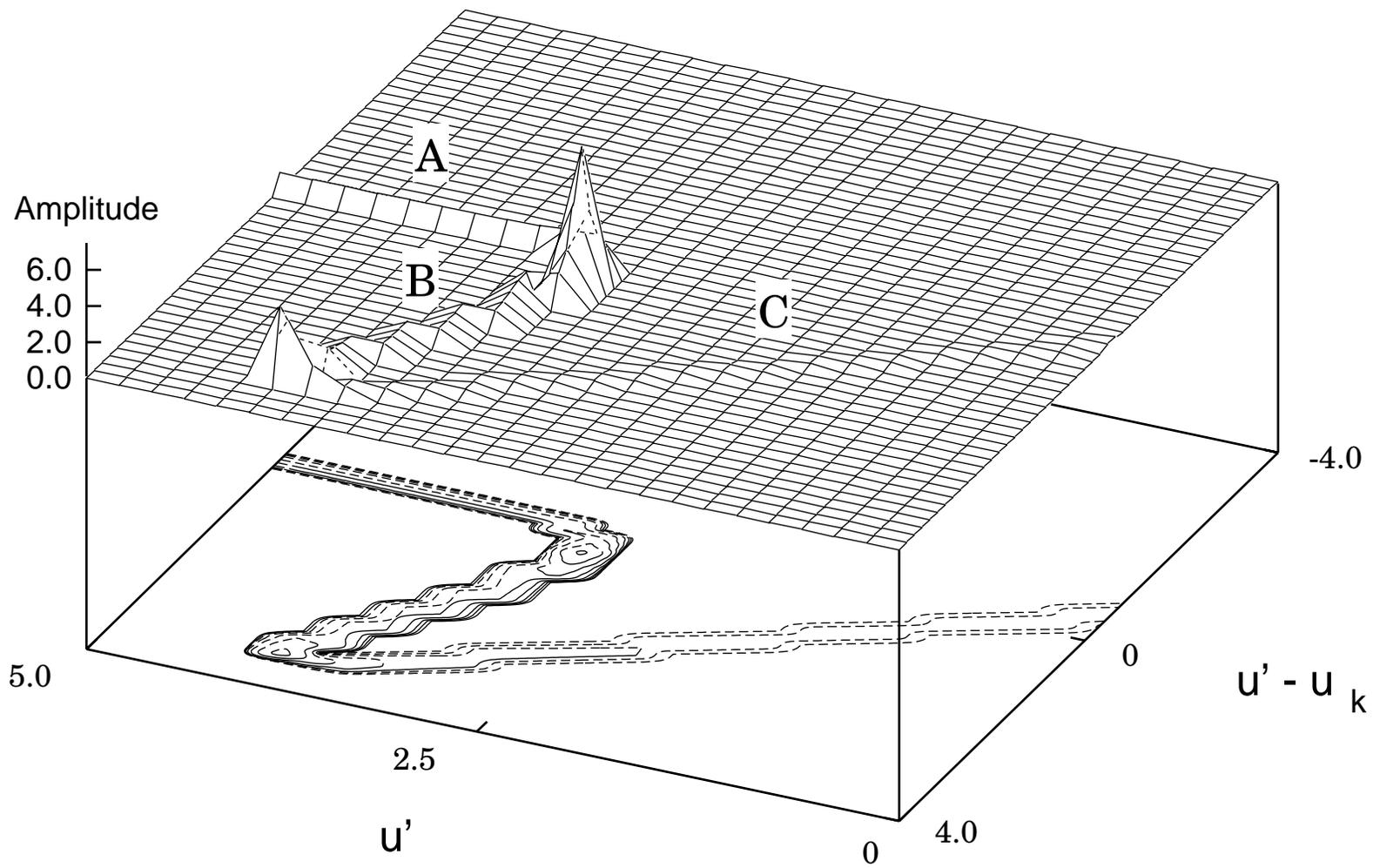
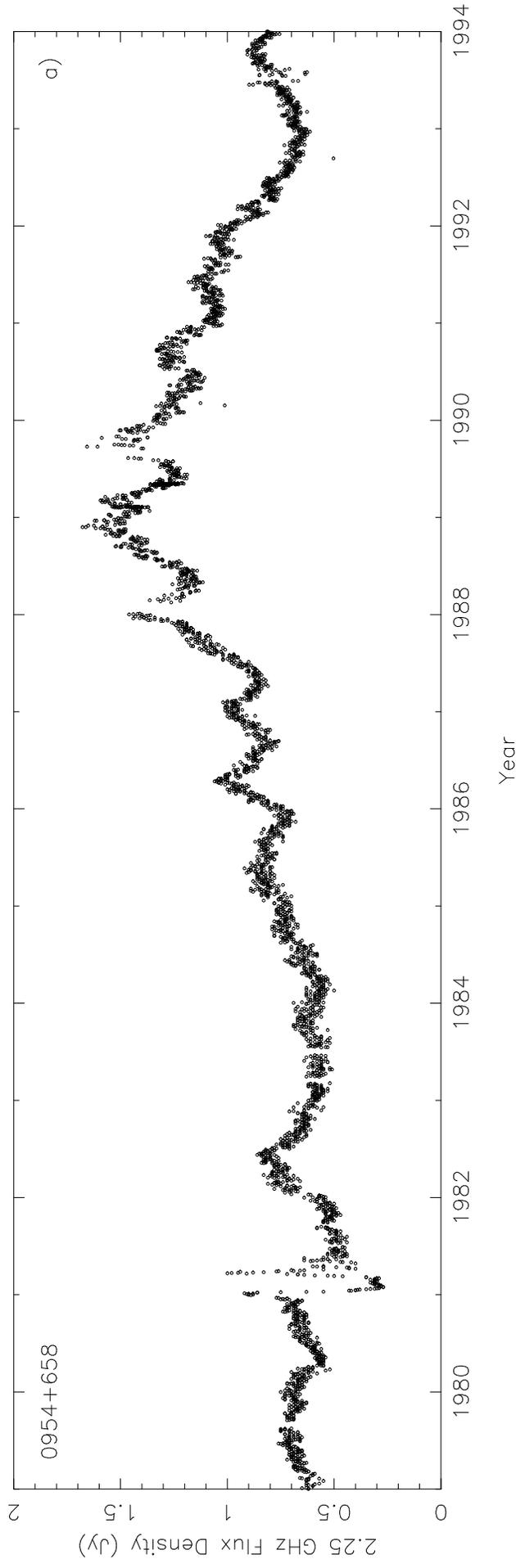
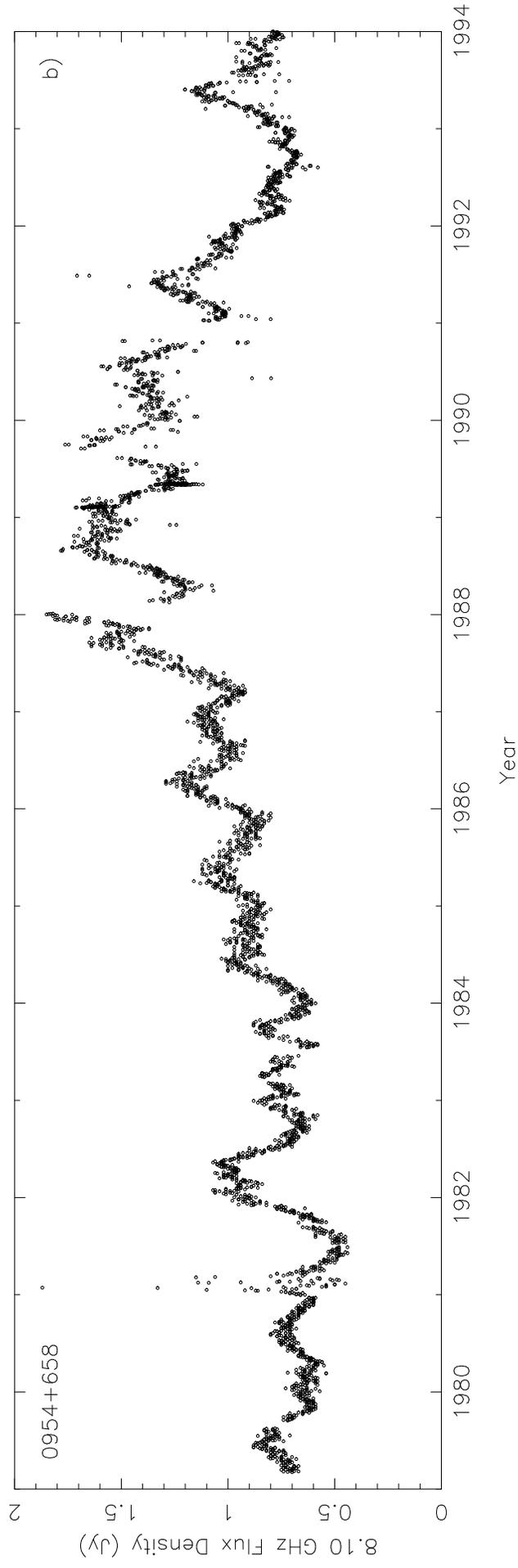
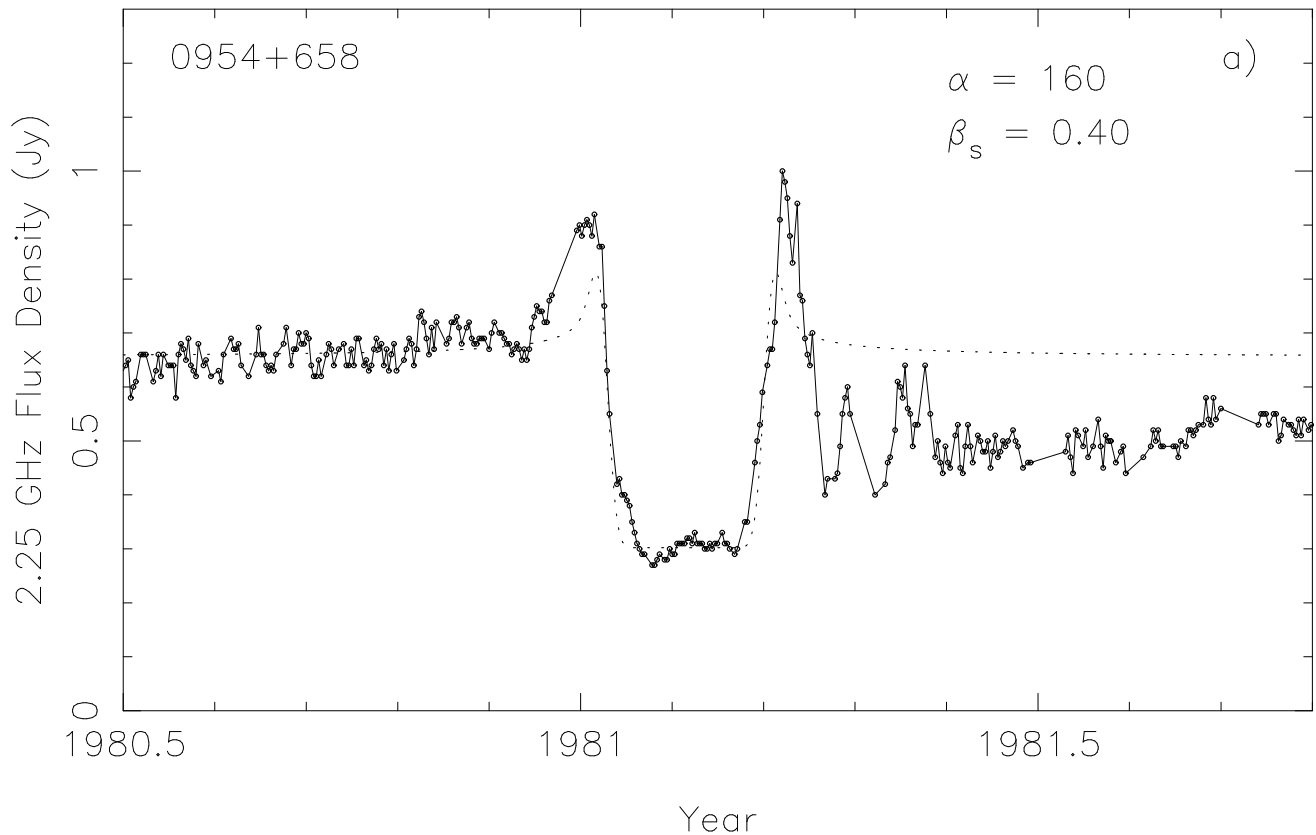
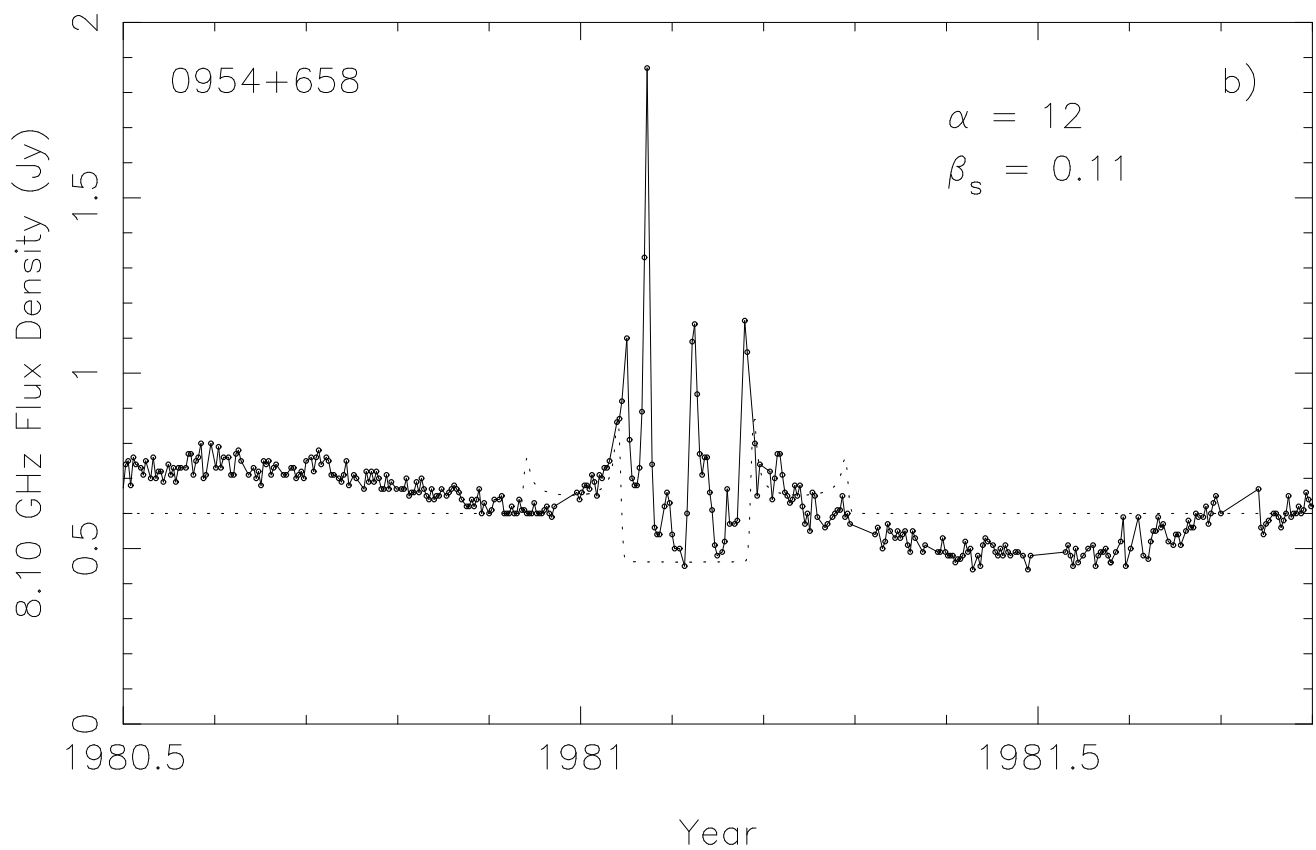
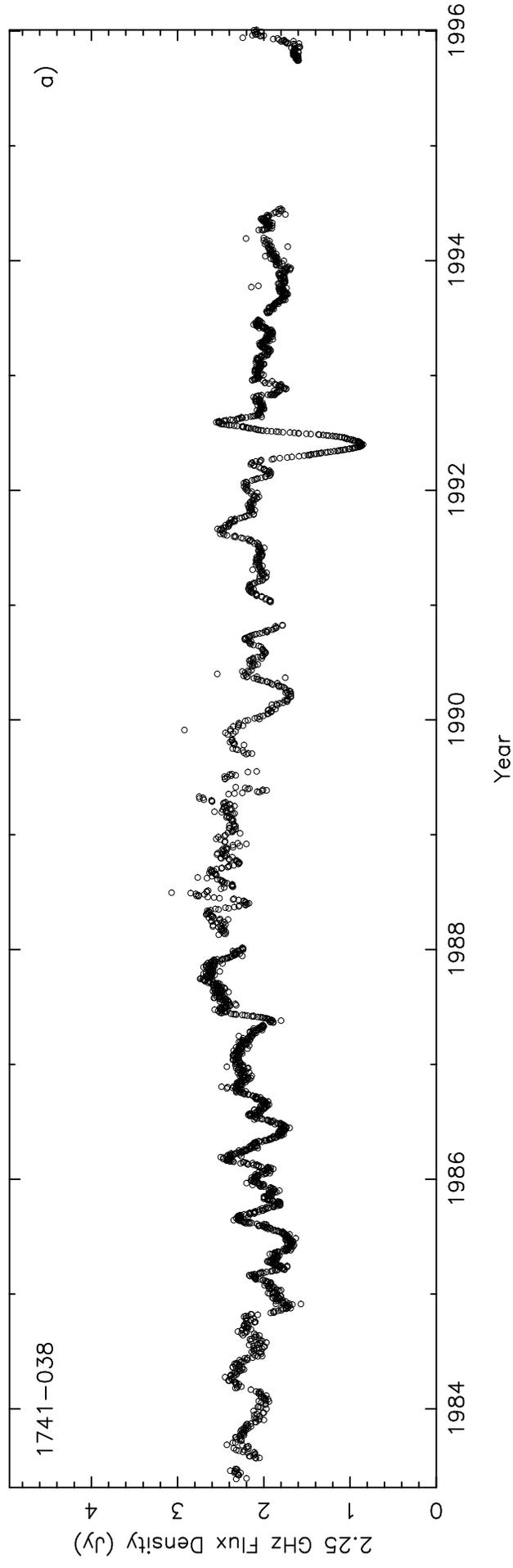
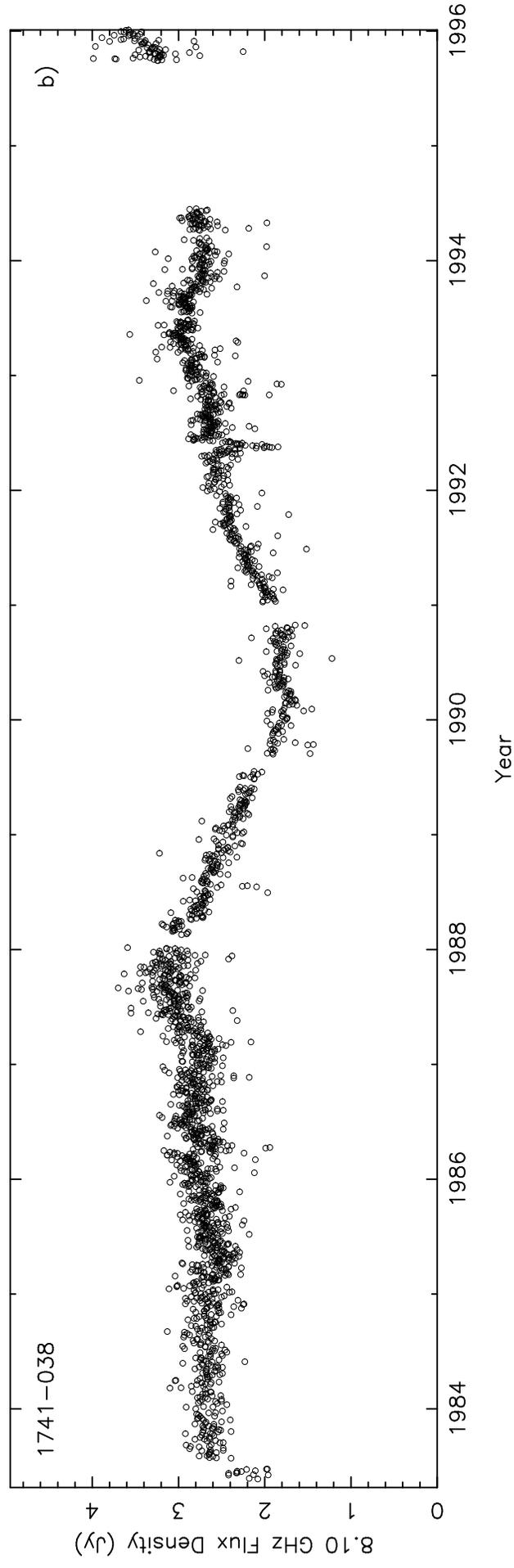
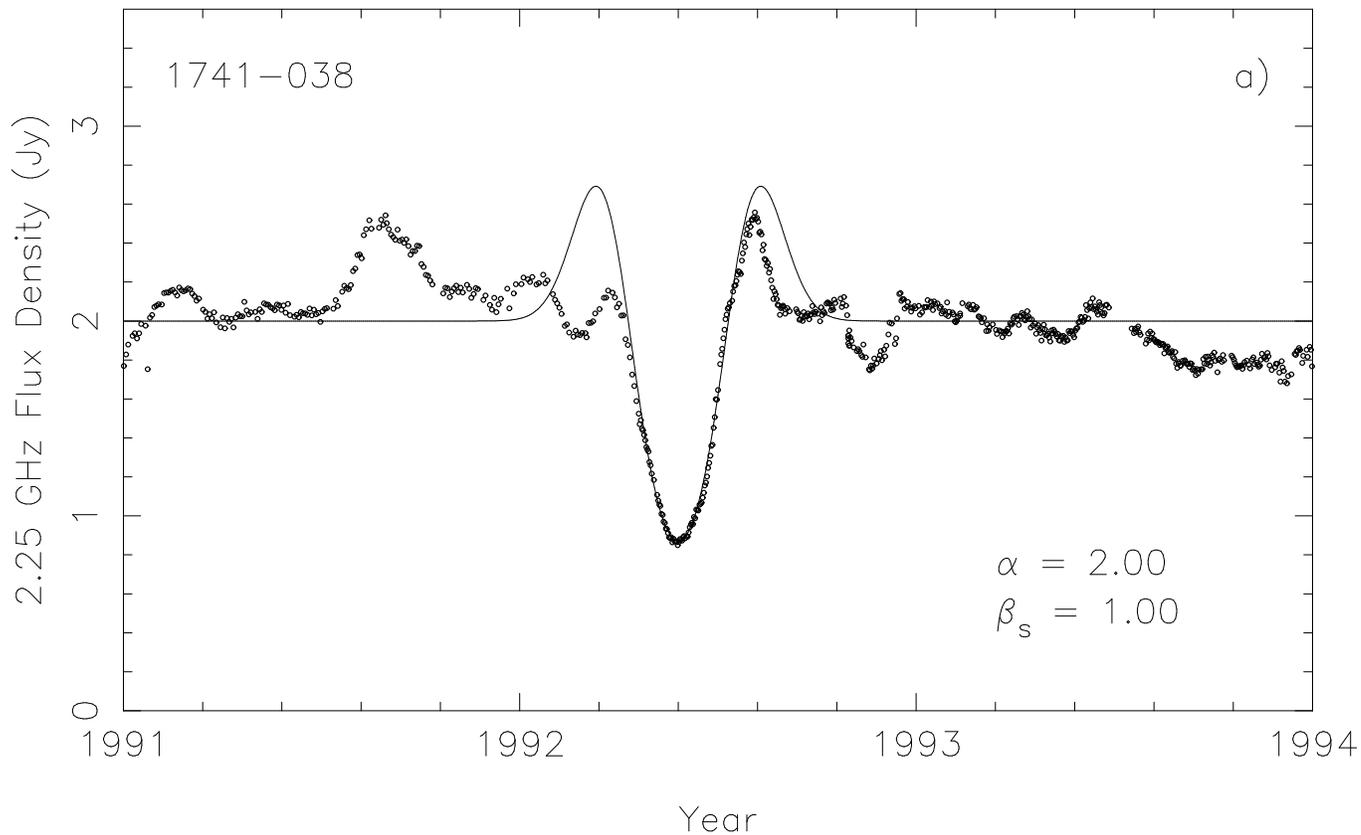
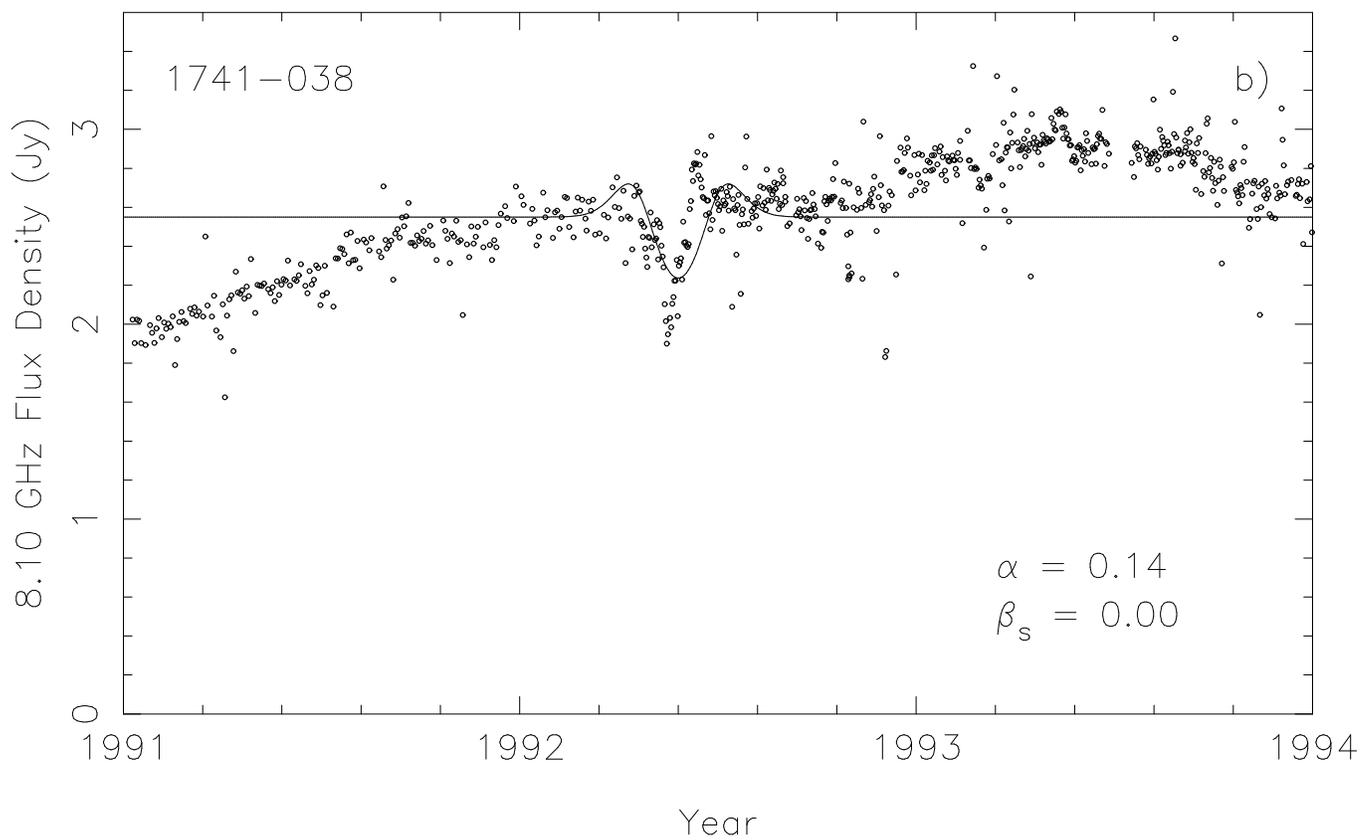
\end{document}